\documentclass[fleqn,usenatbib]{aa}  

\usepackage{graphicx}
\usepackage{txfonts}
\usepackage{multicol}
\usepackage{color,ulem}
\usepackage{amsmath, epsfig,natbib}
\usepackage{color,ulem}
\usepackage{float}
\usepackage{newtxtext,newtxmath}
\definecolor{webgreen}{rgb}{0,.5,0}
\definecolor{webbrown}{rgb}{.6,0,0}
\usepackage[colorlinks=true,hyperfootnotes=false,%
   breaklinks=true,%
   plainpages=false, bookmarksnumbered, bookmarksopen=true,
  bookmarksopenlevel=1,%
   urlcolor=webbrown, linkcolor=blue, citecolor=blue]{hyperref}
\newcommand{\pc}{\>{\rm pc}}
\newcommand{\kpc}{\mbox{$\>{\rm kpc}$}} 

\newcommand{\Gyr}{\mbox{$\>{\rm Gyr}$}}
\newcommand{\Myr}{\mbox{$\>{\rm Myr}$}}

\newcommand\degrees{^\circ}
\newcommand{\avg}[1]{\mbox{$\left<{#1}\right>$}}

\begin{document} 

   \title{Bars and boxy/peanut bulges in thin and thick discs}
   \subtitle{II. Can bars form in hot thick discs?}
\titlerunning{Bar formation in thin and thick disc}
   \author{
   Soumavo Ghosh\inst{1} \thanks{E-mail: ghosh@mpia-hd.mpg.de},
   Francesca Fragkoudi \inst{2},
   Paola Di Matteo \inst{3}
   \and
   Kanak Saha \inst{4}
   }
\authorrunning{S. Ghosh et al.}
   \institute{Max-Planck-Institut f\"{u}r Astronomie, K\"{o}nigstuhl 17, D-69117 Heidelberg, Germany
   \and
   Institute for Computational Cosmology, Department of Physics, Durham University, South Road, Durham DH1 3LE, UK
     \and
    GEPI, Observatoire de Paris, PSL Research University, CNRS, Sorbonne Paris Cité, 5 place Jules Janssen, 92190, Meudon, France
    \and 
    Inter-University Centre for Astronomy and Astrophysics, Pune 411007, India
      }
   
   \date{Received XXX; accepted YYY}

  \abstract
  {
  The Milky Way as well as a majority of external galaxies possess a thick disc.
  However, the dynamical role of the (geometrically) thick disc on the bar formation and evolution is not fully understood. Here, we investigate the effect of thick discs in bar formation and evolution by means of a suite of $N$-body models of (kinematically cold) thin-(kinematically hot) thick discs. We systematically vary the mass fraction of the thick disc, the thin-to-thick disc scale length ratio as well as thick disc's scale height to examine the bar formation under diverse dynamical scenarios. Bars form almost always in our models, even in presence of a massive thick disc.  The part of the bar constituted by the thick disc closely follows the overall growth and temporal evolution of the part of the bar constituted by the thin disc, only the part of the bar in the thick disc is weaker than the part of the bar in the thin disc. The formation of stronger bars is associated with a simultaneous larger loss of angular momentum and a larger radial heating. In addition, we demonstrate a preferential loss of angular momentum and a preferential radial heating of disc stars, along the azimuthal direction within the extent of the bar, in both thin and thick disc stars. For purely thick disc models (without any thin disc), the bar formation critically depends on the disc scale length and scale height. A larger scale length and/or a larger vertical scale height delays the bar formation time and/or suppresses the bar formation almost completely in thick-disc-only models. We find that the Ostriker-Peeble criterion predicts  the bar instability scenarios in our models better than the Efstathiou-Lake-Negroponte criterion. 
  }


   \maketitle
%

\section{Introduction}
\label{sec:Intro}

Stellar bars are one of the most abundant non-axisymmetric structure in disc galaxies. About two-third of the disc galaxies in the local Universe harbour stellar bars, as revealed by various optical and near-infrared surveys of galaxy morphology \citep[e.g., see][]{Eskridgeetal2000,Whyteetal2002,Aguerrietal2009,NairandAbraham2010,Mastersetal2011,Taehyunetal2015,Kruketal2017}, and about one-third of them host strong bars. The bar fraction varies strongly with the stellar mass \citep[e.g.,][]{NairandAbraham2010}, Hubble type \citep[e.g.,][]{Aguerrietal2009,Butaetal2010,NairandAbraham2010,Barwayetal2011} of the host galaxies. The question remains whether the remaining one-third of disc galaxies in the local Universe are hostile to  bar formation and their subsequent growth, or perhaps whether bars are destroyed during their evolutionary pathway?
\par
Much of our current understanding of the bar formation and its growth in disc galaxies are gleaned from numerical simulations. Several studies using $N$-body simulations have shown that an axisymmetric disc galaxy forms a bar spontaneously when the disc becomes unstable to the formation of bar. When a bar forms, orbits that are close to circular become more elongated, with the bar being comprised of elongated orbits \citep[which are called the $x_1$ family of orbits; e.g., see][]{ContopoulosandGrosbol1989,Martinez-Valpuestaetal2006}. Past theoretical studies have shown that a massive central mass concentration and/or inflow of the interstellar gas in the central region can weaken/destroy stellar bars \citep[e.g., see][]{Pfenniger1990,ShenSellwood2004,Athanassoulaetal2005,Bournaudetal2005,HozumiHernqusit2005,Athanassoula2013}. While, in principle, these mechanisms can lead to (complete) destruction of bars, it might still require a very high central mass concentration or prodigious amount of gas inflow \citep{Athanassoulaetal2005}. Furthermore, recent work by \citet{Ghoshetal2021} demonstrated that a minor merger can also lead to a substantial weakening of stellar bars, and in some cases, even a complete destruction of stellar bars in host galaxies.
\par
On the other hand, a disc galaxy might be dynamically hot enough, or equivalently, having a higher value of Toomre $Q$ parameter to prevent the bar instability which in turn, causes a disc galaxy to remain unbarred throughout their life-time \citep[e.g., see][]{Toomre1964,BinneyTremaine2008}. Also, a (rigid) compact and dense dark matter halo was believed to suppress the bar instability in a disc galaxy \citep[e.g.,][]{Mihosetal1997}. Later $N$-body studies with the dark matter halo treated as \textit{live}, revealed the opposite trend, i.e.,  disc galaxies with larger halo concentration develop a much stronger, larger and thinner bar; however, the bar formation is still delayed. \citep[e.g, see][]{DebattistaandSellwood1998, DebattistaandSellwood2000,Athanassoula2002,Athanassoula2003}.
\par
Several theoretical efforts have been made towards determining a dynamical condition for a disc to become bar-unstable. Earlier works of  \citet{OstrikerandPeebles1973} showed that a stellar disc would enter into the bar instability phase if the ratio of the rotational kinetic energy to the potential energy, $W$, exceeds a critical limit of $0.14 \pm 0.003$. This criterion has been tested further in recent $N$-body simulation \citep[e.g., see][]{SahaElmegreen2018}. On the other hand, \citet{Efstathiouetal1982}, using $2$-D $N$-body simulations (with rigid dark matter halo), proposed an analytic criterion for the global stability of cold exponential stellar discs. To illustrate, a  disc becomes bar-stable if the dimensionless quantity $\varepsilon = \frac{V_{\rm max}}{(\alpha M_{\rm disc} G)^{1/2}}$ becomes greater than $1.1$, 
where $V_{\rm max}$ is the maximum rotational velocity, $M_{\rm disc}$ is the total disc mass, and $\alpha \  (= R_{\rm d}^{-1})$ is the inverse of the disc scale length \citep{Efstathiouetal1982}. Later, this criterion has been tested in context of galaxies from the cosmological zoom-in simulations as well as for the observed galaxies \citep[e.g., see recent works of ][]{Izquierdo-Villalbaetal2022,Romeoetal2022}.  
\par
 Bars are present in the high-redshift galaxies as well. However, some studies claimed a decreasing bar fraction with increasing redshift \citep[e.g., see][]{Shethetal2008,Melvinetal2014,Simmonsetal2014}, while some other studies showed a constant bar fraction up to redshift $z \sim$ 1 \citep[e.g.,][]{Elmetal2004,Jogeeetal2004}. Hence,  the fraction of galaxies  in the high-redshift hosting bar still remains debated. In addition, cosmological simulations find that bars start forming already at $z \sim 1$ \citep[e.g., see][]{Kraljicetal2012,Fragkoudietal2020,Fragkoudietal2021,Rosas-Guevaraetal2022}. Furthermore, recent photometric study by \citet{Guoetal2022} using the rest-frame Near-Infrared images from the James Webb Space Telescope (JWST), unveiled the presence of stellar bars in disc galaxies at high redshifts ($z>1$). At high redshift, discs are known to be thick, kinematically hot (and turbulent), and more gas-rich. So, the question remains - can bars form in such thick discs?
\par
The existence of a thick-disc component is now well-established observationally in the Milky Way as well as in external galaxies \citep[e.g., see][]{Tsikoudi1979,Burstein1979,GilmoreandReid1983,Yoachim2006,Comenronetal2011a,Comeron2011b,Comeronetal2018}. For external galaxies, thick discs are found along the whole Hubble sequence, from early-type, S0 galaxies \citep{Pohlenetal2004,Kasparovaetal2016,Comeronetal2016,Pinnaetal2019b,Pinnaetal2019a} to late-type galaxies \citep{Yoachim2006,Comeronetal2019,matigetal2021,Scottetal2021}. The thick disc component is, in general, vertically more extended, and kinematically hotter as compared to the thin disc component \citep[e.g., see][]{Juricetal2008,Bovyetal2012,Bovyetal2016,Vieiraetal2022}. As for the Milky Way, recent studies of chemical evolution of the $\alpha$-enhanced stars indicated that the mass of the  chemically thick disc can be of the same order as the thin disc \citep[e.g.,][]{Haywoodetal2013,Snaithetal2015}. Also, \citet{Comenronetal2011a} argued that (geometrically) thick discs can constitute a significant fraction of the baryonic content of galaxies. 
\par
While significant efforts have been devoted towards understanding the bar instability scenario in disc galaxies, a majority of the $N$-body simulations considered a one-component stellar disc. A few studies in the past have investigated how the disc stars would get trapped in the bar instability as a function of how dynamical hot or cold the underlying population is \citep[e.g., see][]{Hohl1971,AthanassoulaandSellwood1986,Athanassoula2003,Debattistaetal2017}. \citet{AumerandBinney2017} showed that the presence of a thick disc delays the bar formation. In addition, past study by \citet{Klypinetal2009} showed that $N$-body models with thick discs (scale height-to-length ratio = 0.2) produce very long and slowly rotating bars. Furthermore, \citet{Fragkoudietal2017} studied the effect of such a thick disc component on bar and the boxy/peanut formation using a fiducial two-component thin+thick model where the thick disc constitutes 30 percent of the total stellar mass. However, a systematic study of bar formation in discs with different hot and cold discs, as well as composite thin and thick discs is still missing. We aim to address this here.
\par
In this work, we systematically investigate the dynamical role of the thick disc component in bar formation and growth by making use of a suite of $N$-body models with (kinematically cold) thin and (kinematically hot) thick discs. Within the suite of $N$-body models, we vary the thick disc mass fraction as well as consider different geometric configurations (varying ratio of the thin and thick disc scale lengths). Furthermore, for some models, we vary the disc scale height as well (while keeping the scale length fixed) to examine the bar formation scenario in such cases. We quantify the strength and growth of the bars in both the thin and thick disc stars, and also study the underlying dynamical mechanisms, e.g., angular momentum transport and the radial heating within the bar region. In addition, we test some of the most commonly used bar instability criteria, as mentioned before, on the suite of $N$-body models considered here. 
\par
The rest of the paper is organised as follows.
Sect.~\ref{sec:sim_setup} provides the details of the simulation set-up and the initial equilibrium models. Sect.~\ref{sec:bar_evolution_fthick} quantifies the properties of the bars in different models and the associated temporal evolution while sect.~\ref{sec:disc_scale_height} discusses the effect of disc scale height on the bar formation.  Sect.~\ref{sec:angmom_radialHeating_withinbar} provides the details of the angular momentum transport and the radial heating within the bar region. Sect.~\ref{sec:bar_instability} contains the results related to applying a few   instability criteria on the thin+thick disc models considered here. Sect.~\ref{sec:discussion} contains the discussion while  Sect.~\ref{sec:conclusion} summarizes the main findings of this work.

\section{Simulation set-up \& $N$-body models}
\label{sec:sim_setup}
\begin{table*}
\centering
\caption{Key structural parameters for the equilibrium models.}
\begin{tabular}{cccccccccccc}
\hline
\hline
 Model$^{(1)}$ & $f_{\rm thick}$$^{(2)}$ & $R_{\rm d, thin}$$^{(3)}$ & $R_{\rm d, thick}$$^{(4)}$ & &  $z_{\rm {d, thin}}$$^{(5)}$ & $z_{\rm {d, thick}}$$^{(6)}$ & $M_{\rm star}$$^{(7)}$ & $R_{\rm H}$$^{(8)}$ & $M_{\rm dm}$$^{(9)}$ & $n_{\rm star}$$^{(10)}$ & $n_{\rm dm}$$^{(11)}$ \\
\\
 && (kpc) & (kpc) & & (kpc) & (kpc) & ($\times 10^{11} M_{\odot}$) & (kpc) & ($\times 10^{11} M_{\odot}$) & ($\times 10^{5}$) & ($\times 10^{5}$)  \\
\hline
rthick0.0 & -- & 4.7 &  -- &  & 0.3 & -- & 1 & 10 & 1.6 & 10 & 5 \\
\hline
rthickS0.1 & 0.1 & 4.7 &  2.3 &  & 0.3 & 0.9 & 1 & 10 & 1.6 & 10 & 5 \\
rthickE0.1 & 0.1 & 4.7 &  4.7 &  & 0.3 & 0.9 & 1 & 10 & 1.6 & 10 & 5 \\
rthickG0.1 & 0.1 & 4.7 &  5.6 &  & 0.3 & 0.9 & 1 & 10 & 1.6 & 10 & 5 \\
rthickS0.3 & 0.3 & 4.7 &  2.3 &  & 0.3 & 0.9 & 1 & 10 & 1.6 & 10 & 5 \\
rthickE0.3 & 0.3 & 4.7 &  4.7 &  & 0.3 & 0.9 & 1 & 10 & 1.6 & 10 & 5 \\
rthickG0.3 & 0.3 & 4.7 &  5.6 &  & 0.3 & 0.9 & 1 & 10 & 1.6 & 10 & 5 \\
rthickS0.5 & 0.5 & 4.7 &  2.3 &  & 0.3 & 0.9 & 1 & 10 & 1.6 & 10 & 5 \\
rthickE0.5 & 0.5 & 4.7 &  4.7 &  & 0.3 & 0.9 & 1 & 10 & 1.6 & 10 & 5 \\
rthickG0.5 & 0.5 & 4.7 &  5.6 &  & 0.3 & 0.9 & 1 & 10 & 1.6 & 10 & 5 \\
rthickS0.7 & 0.7 & 4.7 &  2.3 &  & 0.3 & 0.9 & 1 & 10 & 1.6 & 10 & 5 \\
rthickE0.7 & 0.7 & 4.7 &  4.7 &  & 0.3 & 0.9 & 1 & 10 & 1.6 & 10 & 5 \\
rthickG0.7 & 0.7 & 4.7 &  5.6 &  & 0.3 & 0.9 & 1 & 10 & 1.6 & 10 & 5 \\
rthickS0.9 & 0.9 & 4.7 &  2.3 &  & 0.3 & 0.9 & 1 & 10 & 1.6 & 10 & 5 \\
rthickE0.9 & 0.9 & 4.7 &  4.7 &  & 0.3 & 0.9 & 1 & 10 & 1.6 & 10 & 5 \\
rthickG0.9 & 0.9 & 4.7 &  5.6 &  & 0.3 & 0.9 & 1 & 10 & 1.6 & 10 & 5 \\
rthickS1.0 & 1 & -- &  2.3 &  & -- & 0.9 & 1 & 10 & 1.6 & 10 & 5 \\
rthickE1.0 & 1 & -- &  4.7 &  & -- & 0.9 & 1 & 10 & 1.6 & 10 & 5 \\
rthickG1.0 & 1 & -- &  5.6 &  & -- & 0.9 & 1 & 10 & 1.6 & 10 & 5 \\
\hline
rthickS1.0$z_d2.3$ & 1 & -- &  2.3 &  & -- & 2.3 & 1 & 10 & 1.6 & 10 & 5 \\
rthickE1.0$z_d2.3$ & 1 & -- &  4.7 &  & -- & 2.3 & 1 & 10 & 1.6 & 10 & 5 \\
rthickG1.0$z_d2.3$ & 1 & -- &  5.6 &  & -- & 2.3 & 1 & 10 & 1.6 & 10 & 5 \\
rthickE1.0$z_d4.7$ & 1 & -- &  4.7 &  & -- & 4.7 & 1 & 10 & 1.6 & 10 & 5 \\
rthickG1.0$z_d4.7$ & 1 & -- &  5.6 &  & -- & 4.7 & 1 & 10 & 1.6 & 10 & 5 \\
rthickG1.0$z_d5.6$ & 1 & -- &  5.6 &  & -- & 5.6 & 1 & 10 & 1.6 & 10 & 5 \\
\hline
\hline
\end{tabular}
{\newline
 (1) - name of the model; (2) - thick-disc mass fraction; (3) - scale length of the thin disc; (4) - scale length of the thick disc; (5) - scale height of the thin disc; (6) - scale height of the thick disc; (7) - mass of the stellar (thin+thick) disc; (8) - characteristic scale length of the dark matter halo; (9) - mass of the dark matter halo; (10) - total number of particles in the stellar (thin+thick) disc; (11) - total number of particles in the dark matter halo. }
\label{table:key_param}
\end{table*}
To motivate our study, we develop a suite of $N$-body models, consisting of a thin and a thick stellar disc, and the whole system is embedded in a live dark matter halo. One such model is already presented in \citet{Fragkoudietal2017}. Here, we build a  suite of numerical models of thin+thick discs while systematically varying the thick disc mass fraction as well as the ratio of the thick-to-thin disc scale lengths.
\par
Each of the thin and thick discs is modelled with a Miyamoto-Nagai profile whose potential has the form \citep{MiyamatoandNagai1975}
\begin{equation}
\Phi_{\rm d} = -\frac{GM_{\rm d}}{\sqrt{R^2+\left(R_{\rm d}+\sqrt{z^2+z_{\rm d}^2}\right)^2}}\,,
\label{eq:pot_disc}
\end{equation}
\noindent where $R_{\rm d}$ and $z_{\rm d}$ are the characteristic disc scale length and scale height, respectively, and  $M_{\rm d}$ is the total mass of the disc. The dark matter halo is modelled with a Plummer sphere whose potential has the form \citep{Plummer1911}
\begin{equation}
\Phi_{\rm dm} (r) = - \frac{GM_{\rm dm}}{\sqrt{r^2+R^2_{\rm H}}}\,,
\label{eq:pot_dm}
\end{equation}
\noindent where $R_{\rm H}$ is the characteristic scale length and $M_{\rm dm}$ is the total halo mass. Here, $r$ and $R$ are the radius in the spherical and the cylindrical coordinates, respectively. The values of the key structural parameters for the thin, thick discs as well as the dark matter halo are listed in Table.~\ref{table:key_param}. The total number of particles used to model each of these structural components are also mentioned in Table.~\ref{table:key_param}.
\par
The initial conditions of the discs are obtained using the `iterative method' algorithm \citep[see][]{Rodionovetal2009}. This algorithm constructs equilibrium phase models for stellar systems by making use of a constrained evolution in such a way that the equilibrium solution has a number of desired parameters. For this work, we only constrain the density profile of stellar discs (following Eq.~\ref{eq:pot_disc}) while allowing the velocity dispersions (both radial and vertical components) to vary in such a way that the system converges to an equilibrium solution \citep[for details, see][]{Fragkoudietal2017}.
\par
In the initial equilibrium set-up for the thin+thick disc models, we consider three different scenarios for the scale lengths of the two disc (thin and thick) components. In \textbf{rthickE} models, both the scale lengths of thin and thick discs are kept same ($R_{\rm d, thick} = R_{\rm d, thin}$) whereas in \textbf{rthickG} models, the scale length of the thick disc component is larger than that for the thin disc ($R_{\rm d, thick} > R_{\rm d, thin}$). In \textbf{rthickS} models, the scale length of the thick disc component is shorter than that for the thin disc ($R_{\rm d, thick} < R_{\rm d, thin}$). 
For each of these three configurations, the fraction of total stellar particles that are in thick disc component ($f_{\rm thick}$) \footnote{Or equivalently the mass fraction in the thick disc as all the disc particles have same mass.} is systematically varied from 0.1 to 0.9 (with a step-size of 0.2). Thus, we have a total of 15 such thin+thick models which we analyse later in this work. In addition, we consider one pure thin disc model ($f_{\rm thick} =0$) and three pure thick disc models ($f_{\rm thick} =1$) to augment this study. Furthermore, to study the effect of disc scale height in bar formation (see Sect.~\ref{sec:disc_scale_height}), we constructed another six thick-disc only models while varying the vertical scale height of the thick disc. The values of the key structural parameters for these six models are also listed in Table.~\ref{table:key_param}. Thus, we analysed a total of 25 $N$-body models for this work.
\par
The simulations are run using a TreeSPH code by \citet{SemelinandCombes2002}. This code has been extensively used to simulate both interacting and merging galaxies as well as isolated galaxies \citep[e.g.,][]{Fragkoudietal2017,Jean-Baptisteetal2017}. A hierarchical tree method \citep{BarnesandHut1986} with a tolerance parameter\footnote{It is the angular size of a group of distant particles, seen from the particle. If the angular size of the group, seen from the particle, is smaller than $\theta$, then the tree code computes the contribution  of the gravitational force (by that group of distant particles) acting on a given particle using the multipole moments of the group mass distribution.} $\theta = 0.7$ is employed to calculate the gravitational force which includes terms up to the quadrupole order in the multipole expansion. A Plummer potential is used to soften the gravitational forces with a softening length $\epsilon = 150 \pc$. The equations of motion are integrated using a leapfrog algorithm \citep{Pressetal1986} with a fixed time step of $\Delta t = 0.25 \Myr$. All the models considered here, are evolved for a total time of $9 \Gyr$.
\par
For consistency, any thin+thick model is referred as a unique string `{\sc [model configuration][thick disc fraction]'}. {\sc [model configuration]} denotes the corresponding thin-to-thick disc scale length configuration, i.e., \textbf{rthickG}, \textbf{rthickE}, or \textbf{rthickS} whereas {\sc [thick disc fraction]} denotes the fraction of the total disc mass that is in the thick disc population. To illustrate, \textbf{rthickG0.3} denotes the model where  the scale length of the thick disc component is larger than that for the thin disc, and 30 percent of the total disc mass is in thick disc component. The pure thin-disc only model is referred as \textbf{rthick0.0} whereas the pure thick-disc only models are referred as \textbf{rthickS1.0}, \textbf{rthickE1.0}, and \textbf{rthickG1.0}. Similarly, the six thick-disc-only models (where we varied the scale height) are  referred as \textbf{rthickE1.0}$z_d2.3$, \textbf{rthickG1.0}$z_d2.3$, \textbf{rthickE1.0}$z_d4.7$, \textbf{rthickE1.0}$z_d4.7$ and \textbf{rthickG1.0}$z_d5.6$. To illustrate, \textbf{rthickS1.0}$z_d2.3$ denotes a thick-disc only model with the thick disc scale height set to $2.3 \kpc$ (while keeping other parameters unchanged) and so on.

\section{Bar formation \& evolution  for different thick-disk mass fractions}
\label{sec:bar_evolution_fthick}
%
\begin{figure*}
\centering
\resizebox{0.87\linewidth}{!}{\includegraphics{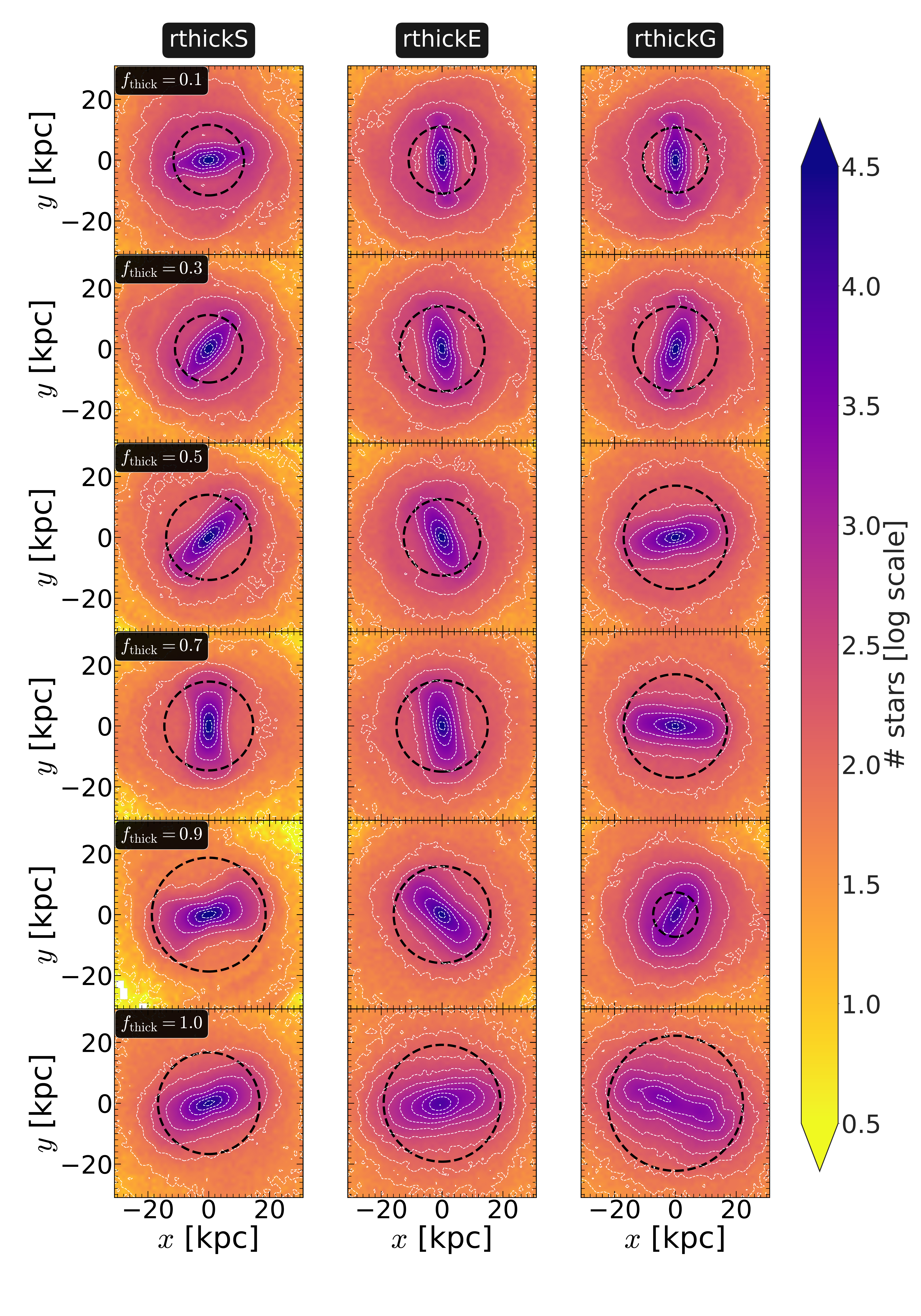}}
\caption{Face-on distribution of all disc particles (thin+thick), at the end of the simulation run ($t = 9 \Gyr$) for all thin+thick disc models with varying $f_{\rm thick}$ values. White solid lines denote the contours of constant density. The black solid circle in each sub-panel denotes the corresponding extent of the bar ($R_{\rm bar}$). \textit{Left-hand panels} show the density distribution for the \textbf{rthickS} models whereas  \textit{middle panels} and \textit{right-hand panels} show the density distribution for the \textbf{rthickE}  and \textbf{rthickG} models, respectively. The thick disc fraction ($f_{\rm thick}$) varies from 0.1 to 1 (top to bottom), as indicated in the left-most panel of each row. A prominent bar forms in almost all models considered here.}
\label{fig:density_maps_endstep_allmodels}
\end{figure*}
%
\begin{figure*}
\centering
\resizebox{0.95\linewidth}{!}{\includegraphics{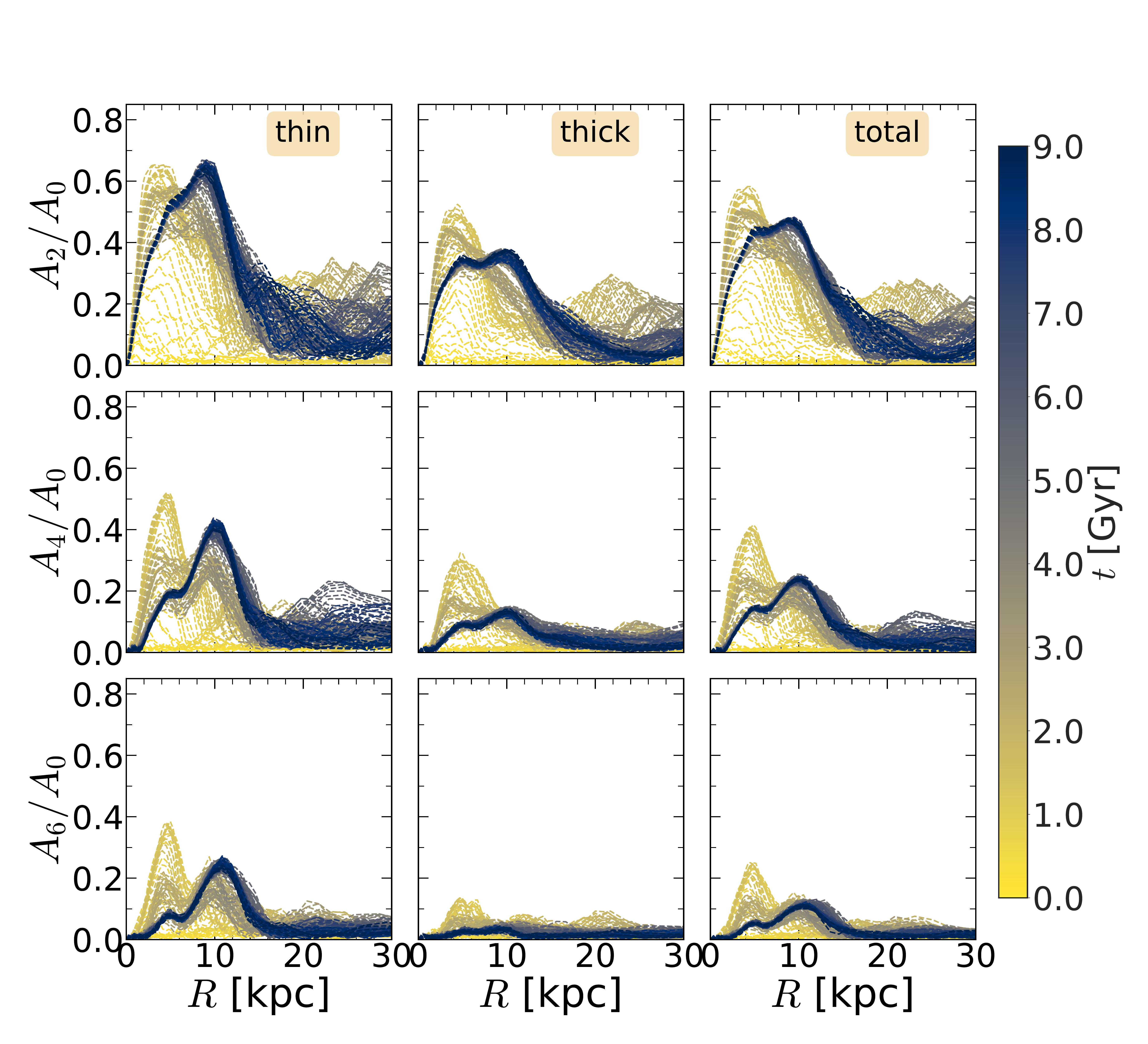}}
\caption{Radial profiles of the $m=2, 4, 6$ Fourier coefficients (normalised by the $m=0$ component), for the thin disc (left-hand panels), thick disc stars (middle panels), and total (thin+thick) disc stars (right-hand panels) as a function of time (shown in colour bar) for the model \textbf{rthickE0.5}. At later times, the peaks of the radial profiles of the $m=2, 4, 6$ Fourier coefficients shift towards the outer regions.}
\label{fig:fourier_profiles_rthickE05}
\end{figure*}
%
%
\begin{figure*}
\centering
\resizebox{\linewidth}{!}{\includegraphics{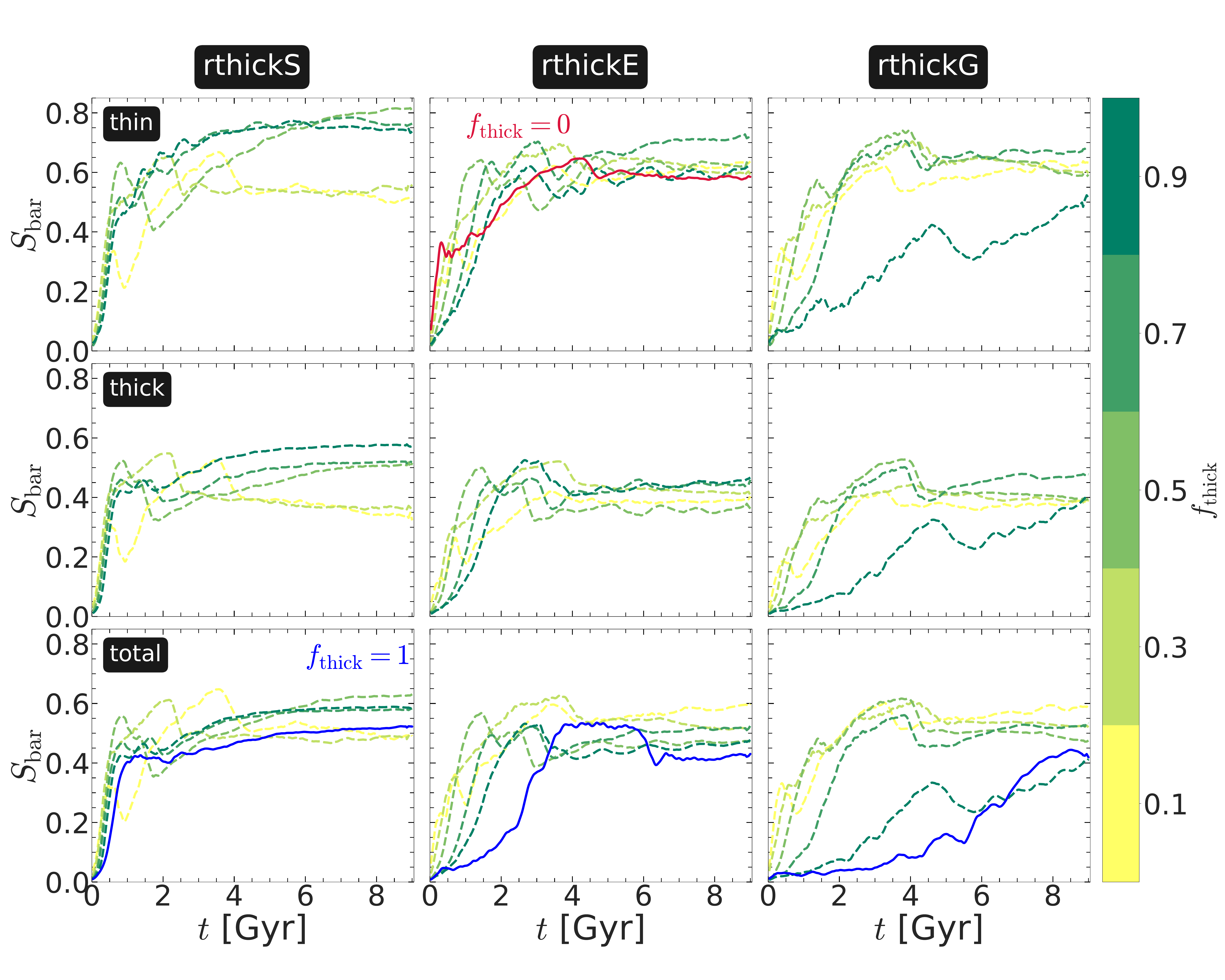}}
\caption{Temporal evolution of the bar strength, $S_{\rm bar}$, for the thin disc (upper panels), thick disc (middle panels), and total (thin+thick) disc stars (lower panels) for all thin+thick disc models with varying $f_{\rm thick}$ values (see the colour bar). \textit{Left-hand panels} show the bar strength evolution for the \textbf{rthickS} models whereas  \textit{middle panels} and \textit{right-hand panels} show the bar strength evolution for the \textbf{rthickE}  and \textbf{rthickG} models, respectively. The thick disc fraction ($f_{\rm thick}$) varies from 0.1 to 0.9 (with a step-size of 0.2), as indicated in the colour bar. The blue solid lines in the bottom row denote the three thick disc-only models ($f_{\rm thick} =1$) whereas the red solid line in the top middle panel denotes the thin disc-only model ($f_{\rm thick} =0$), for details see text.}
\label{fig:barStrength_evolution}
\end{figure*}
\begin{figure*}
\centering
\resizebox{0.9\linewidth}{!}{\includegraphics{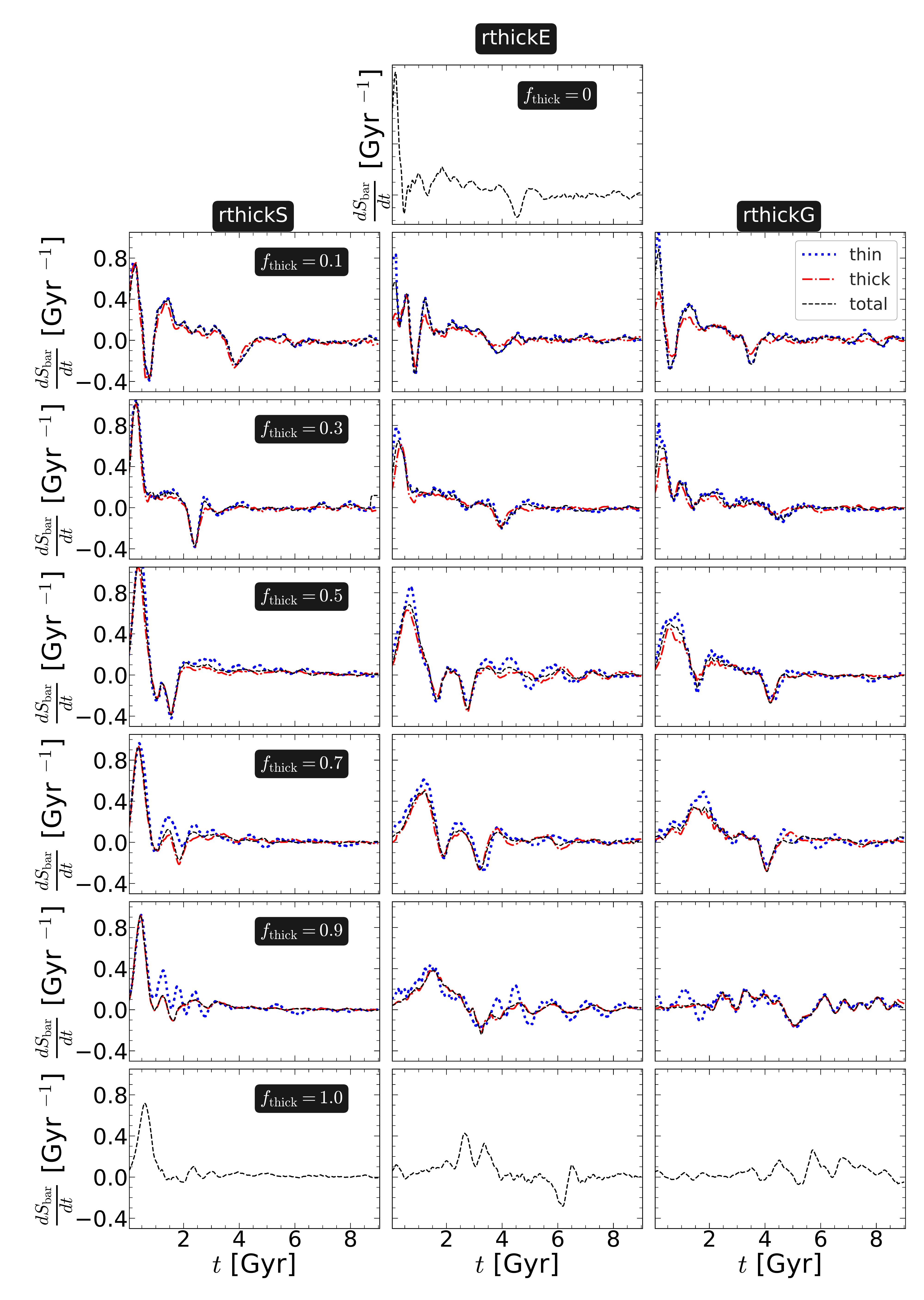}}
\caption{Temporal evolution of the growth/decay rate of the bar strength ($d S_{\rm bar}/dt$), for the thin, thick, and total (thin+thick) disc components, for all models with varying $f_{\rm thick}$ values. \textit{Left-hand panels} show the evolution for the \textbf{rthickS} models whereas  \textit{middle panels} and \textit{right-hand panels} show the evolution for the \textbf{rthickE}  and \textbf{rthickG} models, respectively. The $f_{\rm thick}$ values are indicated in the left-most panel of each row. The top middle row shows the  growth/decay rate of the bar strength for the thin-disc-only model ($f_{\rm thick} =0$).}
\label{fig:barStrength_gradient}
\end{figure*}
%
Before we present the results, we mention that in our thin+thick models, we can identify and separate, by construction, which
stars are members of the thin disc component and which stars are members of the thick disc component as well as we can track them as the system evolves self-consistently. Thus, throughout this paper, we refer to the bar as seen exclusively in the thin disc population as the `thin disc bar' or equivalently `bar in the thin disc' and that seen exclusively in the thick disc population as the `thick disc bar' or equivalently `bar in the thick disc'.

Fig.~\ref{fig:density_maps_endstep_allmodels} shows the density distribution of all stars (thin+thick) in the face-on projection ($x-y$-plane) for all thin+thick disc models (with $f_{\rm thick}$ varying from 0.1 to 1) considered here, at the end of the simulation run ($t = 9 \Gyr$). Even a mere visual inspection reveals that almost all the models develop a strong bar by $9 \Gyr$. We further checked the same face-on density distribution, computed separately for thin and thick disc stars. Both of them show a prominent bar. For the sake of brevity, we have not shown it here \citep[however, see Fig.~2 in][]{Fragkoudietal2017}. 

\par
To quantify the strength of bars in thin and thick disc components of our models, we first compute the radial profiles of the $m=2, 4, 6$ Fourier coefficients using \citep[e.g., see][]{SahaElmegreen2018,Ghoshetal2021}
\begin{equation}
A_m/A_0 (R)= \frac{\sum_i m_i e^{im\phi_i}}{\sum_i m_i}\,,
\label{eq:fourier_calc}
\end{equation}
\noindent where $A_m$ is the coefficient of the $m^{th}$ Fourier moment of the density distribution, $m_i$ is the mass of the $i^{th}$ particle and $\phi_i$ is its cylindrical angle. The summation runs over all the particles within the radial  annulus $[R, R+\Delta R]$, with $\Delta R = 0.5 \kpc$. In Fig~\ref{fig:fourier_profiles_rthickE05}, we show the  corresponding radial profiles of $m=2, 4, 6$ Fourier coefficients for the thin, thick, and thin+thick discs, as a function of time, for the model \textbf{rthickE0.5}. A prominent peak in the $A_2/A_0$ radial profiles, accompanied by a similar peak (with albeit smaller peak value) in the $A_4/A_0$ radial profiles clearly confirm the presence of a strong, central stellar bar in both thin and thick disc components. Moreover, the peaks of the radial profiles of the $m=2, 4, 6$ Fourier coefficients shift towards the outer regions at later times, indicating that the peak of the non-axisymmetry moves towards larger radii as time progresses. This trend is present in both thin and thick disc components.  Furthermore, at later times, the radial profiles of the $m=2$ Fourier coefficient show multiple peaks that are spatially well-separated. We checked that the outward shifting of the peaks in the  radial profiles of the Fourier coefficients as well as the presence of multiple peaks in the $m=2$ profiles are seen in other thin+thick disc models as well. We mention that these models develop a prominent boxy/peanut structure at later times \citep[for details, see][]{Fragkoudietal2017}. The formation of multiple-peaks might be linked to the formation of a 3-D boxy/peanut structure, as demonstrated in \citet{Sahaetal2018} and \citet{Vynatheyaetal2021}. The details of the boxy/peanut formation and their properties is beyond the scope of this paper and will be taken up in a future work.
%
%
\par
Next, we study the temporal evolution of the strength of bars in thin and thick discs separately, as well as for the thin and thick disc combined. At any time $t$, we define the strength of the bar, $S_{\rm bar}$, as the peak value of the $m=2$ Fourier coefficient ($A_2/A_0$). The resulting temporal variation of the bar strength of the thin, thick discs, and thin+thick disc are shown in Fig~\ref{fig:barStrength_evolution} for all models considered here. As seen clearly from Fig.~\ref{fig:barStrength_evolution}, irrespective of the geometric configuration, i.e., the scale length of the thick disc with respect to the scale length of the thin disc, the temporal evolution of the bar strength in the thick disc component closely follow that for the bar in the thin disc component. However, the bar in the thick disc component remains weaker than the bar in the thin disc component; this remains true for all time and for all the thin+thick models considered here. This is in agreement with the earlier findings of \citet{Fragkoudietal2017}. The temporal evolution of $S_{\rm bar}$ for the model \textbf{rthickG0.9} merits some discussion. The bar strength for model \textbf{rthickG0.9} is weaker than for the other \textbf{rthickG} models, as revealed by the values of $S_{\rm bar}$.  
\par
The thin-disc only (\textbf{rthick0.0}) model develops a prominent bar too during the evolution, similar to other thin+thick models. The bar forms quite early (within $\sim 2 \Gyr$), grows for a certain time, and then it saturates (see Fig.~\ref{fig:barStrength_evolution}).  For the thick-disc only models, all three models, namely, \textbf{rthickS1.0}, \textbf{rthickE1.0}, and \textbf{rthickG1.0} form a prominent bar by the end of the simulation run. However, the bar in the \textbf{rthickG1.0} remains weaker when compared with the other two thick-disc-only models (see Fig.~\ref{fig:barStrength_evolution}).
\par
To further quantify the epoch of bar formation and how fast the bar grows in an individual model, we define the growth rate, $d S_{\rm bar}/dt$, as the time derivative of the strength of bar. The resulting variation of the growth rate as a function of time is shown in Fig.~\ref{fig:barStrength_gradient} for thin, thick, and thin+thick disc particles, for all models. In both thin and thick discs, the bar displays a fast growth phase at the initial times, followed by one/multiple buckling phases \citep[for details of buckling instability, see e.g.,][]{Combesetal1990, Martinez-Valpuestaetal2006}, as indicated by the prominent dip (negative values) in the temporal evolution of $d S_{\rm bar}/dt$. However, at later times ($t > 6 \Gyr$), the bars no longer grow as indicated by $d S_{\rm bar}/dt \sim 0$, and remain as a steady non-axisymmetric feature. Also, we mention that we did not find any time-lag between the epochs of bar formation (or bar growth) in the thin and thick discs.
Furthermore, a careful inspection of  Figs.~\ref{fig:barStrength_evolution} and \ref{fig:barStrength_gradient} reveal that the bars in most of the \textbf{rthickS} configuration form at an earlier epoch as compared to other two configurations considered here. Also, during the initial rapid bar formation phase, the bars in the \textbf{rthickS} configuration grows at a much faster rate than the bars in other two configurations, and this trend is seen to be true for almost all $f_{\rm thick}$ values considered here. 
\par
In addition, the temporal evolution of $d S_{\rm bar}/dt$ for the three thick-disc-only models (see bottom panels of Fig.~\ref{fig:barStrength_gradient}) reveals that the bar in the \textbf{rthickG1.0} model forms at a much later time ($\sim 6.1 \Gyr$) while the bar in the \textbf{rthickS1.0} model forms at a much earlier time (within $\sim 1 \Gyr$). Thus, for the \textbf{rthickS} models, irrespective of the mass of the thick disc, the bar seems to grow fast and become quite strong. The \textbf{rthickS} model with 90 percent of the mass in the thick disc still forms a bar. This is also true for the case with 100 percent of the mass in the thick disc. This demonstrates that centrally concentrated discs, even when they are kinematically hot, are unstable to bar formation. As the disc scale length increases (in \textbf{rthickE} and \textbf{rthickG} models), the bar formation epoch gets progressively delayed and also results in forming a progressively weaker bar.
%

\section{Effect of disc scale height on bar formation}
\label{sec:disc_scale_height}

\begin{figure}
\centering
\resizebox{1.05\linewidth}{!}{\includegraphics{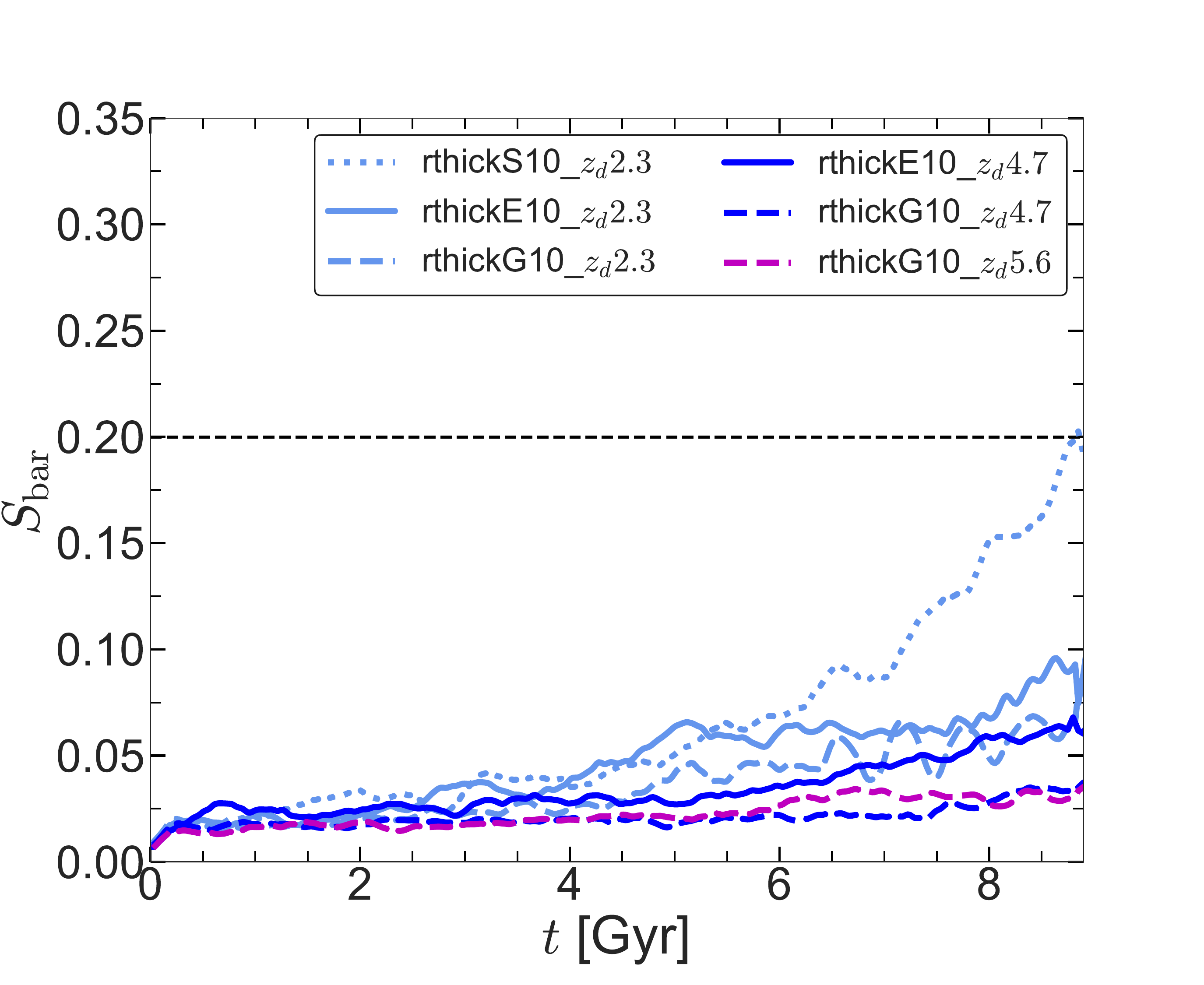}}
\caption{Temporal evolution of the bar strength, $S_{\rm bar}$ for the six thick-disc only models with varying vertical scale height, for details see text. The solid line denotes the \textbf{rthickE} models while the dashed line denotes the \textbf{rthickG} models and the dotted line denotes the \textbf{rthickS} models. Different line colours denote models with different scale-height values. With increasing vertical scale height, the models progressively become bar stable. The horizontal line denotes $S_{\rm bar} =0.2$, and is used as an (operational) limit for the onset of bar formation.}
\label{fig:sbar_new6models}
\end{figure}

Here, we examine the effect of increasing the disc scale height and therefore making it kinematically hotter (in $v_z$) on bar formation for the thick-disc only models. To achieve that, we investigate the growth and evolution of the bar strengths in the six thick-disc-only models where we varied the disc scale height (for details, see Sect.~\ref{sec:sim_setup}).
\par
We calculate the temporal evolution of the bar strength, $S_{\rm bar}$ for these six  models. This is shown in Fig.~\ref{fig:sbar_new6models}. Now, we consider $S_{\rm bar} =0.2$ as the (operational) definition for denoting the onset of the bar, as commonly done in past literature \citep[see e.g.][and references therein]{Ghoshetal2021}. When applied this criterion, we find that  the model \textbf{rthickS1.0}$z_d2.3$ develops a weak bar (with $S_{\rm bar} \sim 0.2$) at around $9 \Gyr$. For the other five models, the $S_{\rm bar}$ values lie below $0.1$ or so, thereby confirming that these model do not form any central bar in $9 \Gyr$. However, for some these models, the temporal evolution of $S_{\rm bar}$ do show an increase (with much swallower slope $d S_{\rm bar}/dt$), thereby indicating they might form a bar at later times ($> 9 \Gyr$).
\par
To conclude, the bar formation gets delayed in presence of a much more vertically-extended thick disc. This is a drastically different scenario when compared to the models \textbf{rthickS1.0} and \textbf{rthickG1.0} which form a prominent bar by the end of the simulation (compare Fig.~\ref{fig:barStrength_evolution} and \ref{fig:sbar_new6models}) in spite of having all the stellar population in the thick disc component (i.e., $f_{\rm thick =1}$). We calculated the radial profiles of the velocity dispersion in the radial as well as in the vertical directions for these six models. We find that as the scale height of the disc is increased, the corresponding vertical velocity dispersion also increased monotonically \citep[see e.g.,][]{vanderKruit2011} while the radial velocity dispersion remains almost unchanged. For the sake of brevity, they  are not shown here. This demonstrates the vital dynamical effect of the vertical scale height and hot vertical kinematics in context of bar formation scenario, in agreement with past findings \citep[e.g., see][]{Combesetal1990,Athanassoula2003}. A similar scenario of suppression of non-axisymmetric instability with increasing disc scale height is also seen for spiral arms \citep{GhoshJog2018,GhoshJog2021}.
%
%

\section{Angular momentum exchange \& radial heating within the bar region}
\label{sec:angmom_radialHeating_withinbar}

It is known that a stellar bar grows in a disc galaxy by transferring angular momentum from the disc to the dark matter halo; this transfer happens preferentially at the bar resonance points \citep[e.g., see][]{TremaineWeinberg1984,HernquistWeinberg1992,DebattistaandSellwood2000,Athanassoula2002,SellwoodandDebattista2006,Dubinskietal2009,SahaNaab2013}. Conversely, the weakening/destruction of a bar is associated with the gain of the angular momentum in the central region of disc galaxies \citep[e.g., see][]{Bournaudetal2005,Ghoshetal2021}. At the same time, as a bar grows, it heats up the disc due to its action on the stars \citep[e.g., see][]{Sahaetal2010,Saha2014}. Here, we quantify the change in the angular momentum content and the radial heating within the bar region as a function of time, for all models (with varying $f_{\rm thick}$ values) considered here.

Before we present the results related to the angular momentum transport and radial heating, we mention that for these analyses we have considered `all' stellar particles that are momentarily within the extent of the bar (defined by $R_{\rm bar}$) at a certain time $t$. In other words, we did not distinguish the stars which are actually part of the bar and which are only stars located within the bar region. The same scheme is applied throughout Sect.~\ref{sec:angmom_radialHeating_withinbar} and in subsequent sections, unless defined otherwise.

\subsection{Exchange of angular momentum}
\label{sec:angmom_exchange}

\begin{figure*}
\centering
\resizebox{0.85\linewidth}{!}{\includegraphics{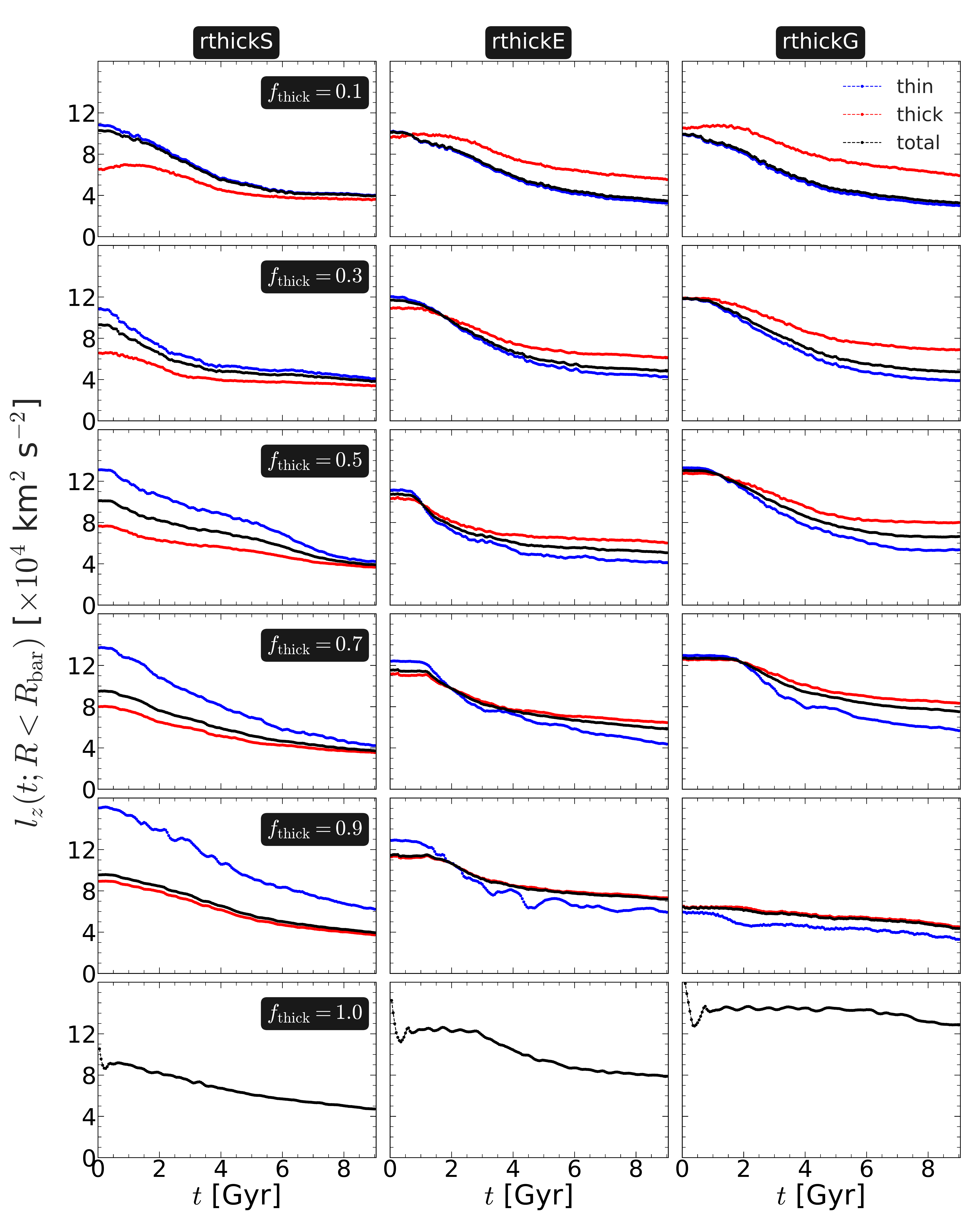}}
\caption{Temporal evolution of the $z$-component of the specific angular momentum, calculated within the extent of the bar, $l_z(t; R < R_{\rm bar})$ (see Eq.~\ref{eq:tempEvol_specficangMom}), for the thin (in blue), thick (in red), and total (thin+thick) disc (in black) particles, for all thin+thick disc models. For each model, the $R_{\rm bar}$ was first fixed to a saturation/representative value, for details see text in section.~\ref{sec:angmom_exchange}. \textit{Left-hand panels} show the evolution for the \textbf{rthickS} models whereas  \textit{middle panels} and \textit{right-hand panels} show the evolution for the \textbf{rthickE}  and \textbf{rthickG} models, respectively. The thick disc fraction ($f_{\rm thick}$) varies from 0.1 to 1 (top to bottom), as indicated in the left-most panel of each row. The extents of bars, used in each sub-panel, are already indicated in Fig~\ref{fig:density_maps_endstep_allmodels} (see black solid circles there).}
\label{fig:lz_evolution_withinbar}
\end{figure*}

At time $t$, we calculate the $z$-component of the angular momentum ($L_z$) of the disc particles, within the bar region, using
\begin{equation}
L_z (t; R < R_{\rm bar}) = \sum_{i=1}^{N(t)} m_i \left[x_i(t) v_{y_i}(t)- y_i(t) v_{x_i}(t)\right]\,,
\label{eq:TempEvo_Lz}
\end{equation}
\noindent where $N(t)$ is the total number of stellar particles contained within the bar region at time $t$, and $x, y, v_x, v_y$ are the position and velocity of the particles. However, we point out that, in our thin+thick models, the fraction of particles assigned to thin and thick discs changes in different model as the thick disc fraction varies across models. Therefore, comparing  $L_z (t; R < R_{\rm bar})$  for various models (with different $f_{\rm thick}$ values) will introduce certain biases as these quantities involved summation over all thin/thick disc particles within the bar region. So, for uniform comparison among all the thin+thick models considered here, we compute the specific angular momentum ($l_z$), within the bar region, using
\begin{equation}
l_z (t; R < R_{\rm bar}) = \frac{1}{N(t)}\sum_{i=1}^{N(t)} \left[x_i(t) v_{y_i}(t)- y_i(t) v_{x_i}(t)\right]\,.
\label{eq:tempEvol_specficangMom}
\end{equation}
The resulting temporal evolution of $l_z$ within the bar region, for the thin, thick, and total (thin+thick) disc particles, are shown in Fig.~\ref{fig:lz_evolution_withinbar} for all thin+thick disc models. We mention that during the total evolution time-span, the bar, in both thin and thick discs, grows substantially (by a factor of $\sim 2$). Therefore, allowing the $R_{\rm bar}$ to vary and then computing the specific angular momentum ($l_z$) within the bar region (using Eq.~\ref{eq:tempEvol_specficangMom}), will essentially blur the intrinsic changes in the specific angular momentum. However, in an individual model, after $6 \Gyr$ or so, the bar does not grow appreciably (see Fig.~\ref{fig:barStrength_evolution}), and the values of $R_{\rm bar}$ also saturates, with occasional fluctuations. Therefore, we take this value as the representative value for the $R_{\rm bar}$ for that particular model. One such $R_{\rm bar}$ determination for the model \textbf{rthickE0.5} is further illustrated in Appendix~\ref{appen:rbar_measurement}.
\par
As seen clearly from Fig.~\ref{fig:lz_evolution_withinbar}, there is a substantial loss of specific angular momentum (within the central bar region, defined by the extent of $R_{\rm bar}$) during the entire evolution of the bar. 
 This trend remains true for both the thin and thick disc components, and for all the thin+thick models. The net loss in the specific angular momentum for the thin disc component is seen to be always larger when compared with that for the thick disc component \citep[as also discussed in][]{Fragkoudietal2017}. At the initial rapid bar growth phase, the loss in the $l_z$ component is larger; however, at later times ($t > 6 \Gyr$ or so) when the bar remains steady, the  loss in the $l_z$ component becomes negligible, and this is true for both the thin and thick disc particles, and for all the thin+thick models considered here. For the thick-disc only models, the loss of the specific angular momentum (within the bar region) is progressively less from \textbf{rthickS1.0} model to \textbf{rthickE1.0} model to \textbf{rthickG1.0} model (see bottom panels of Fig.~\ref{fig:lz_evolution_withinbar}). This matches with the expectation as the formation of progressively weaker bars should be associated with progressively less angular momentum transfer from the central bar region.
\par
%
\begin{figure*}
\centering
\resizebox{0.89\linewidth}{!}{\includegraphics{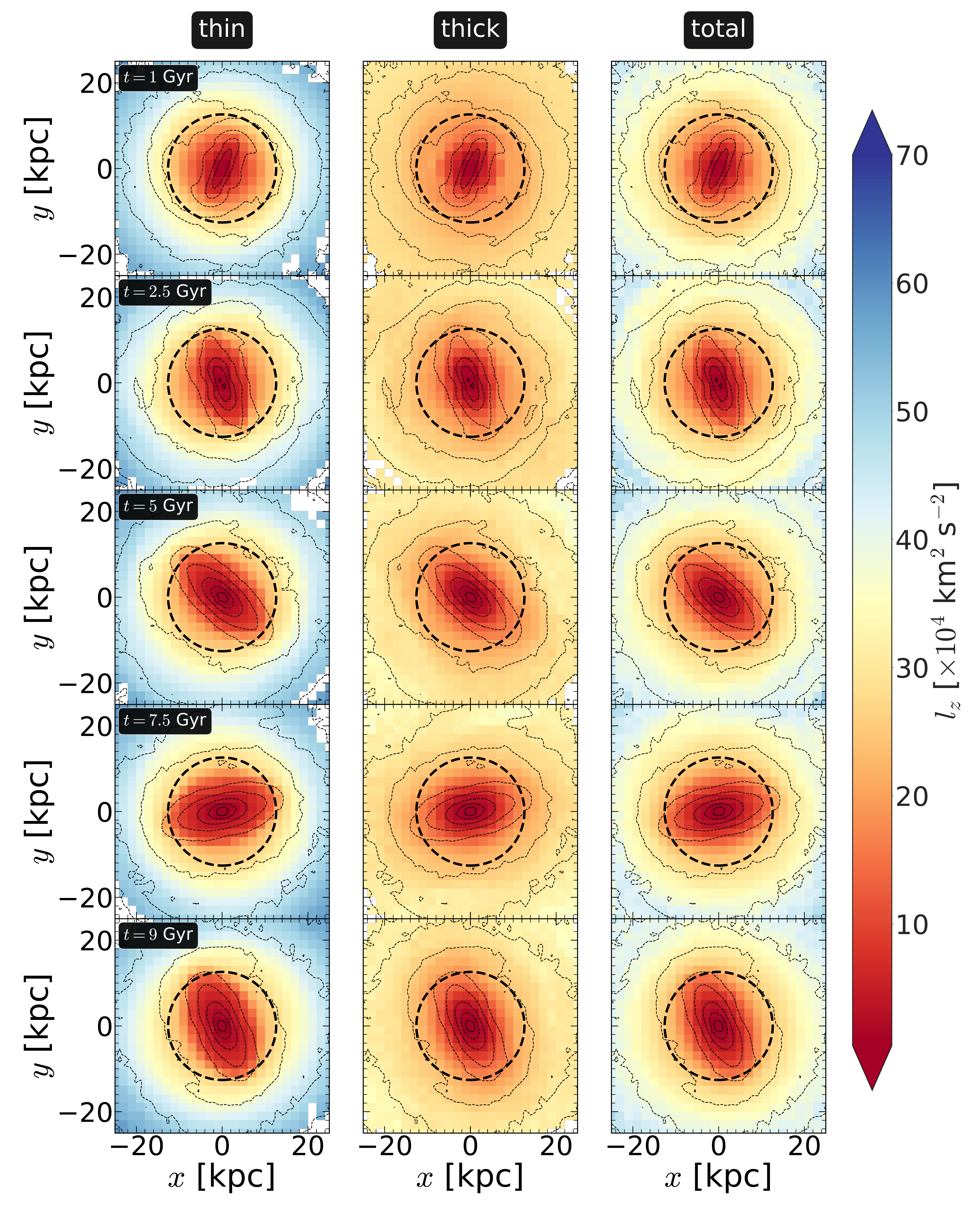}}
\caption{Distribution of the the $z$-component of the specific angular momentum in the face-on  projection (($x,y$)-plane), at different times for the thin (left-hand panels), thick (middle panels), and thin+thick disc particles (right-hand panels) for the model \textbf{rthickE0.5}. Dotted lines denote the contours of total (thin+thick) surface density. The black dashed circle in each sub-panel denote the extent of the bar.} 
\label{fig:speclz_2Dmap}
\end{figure*}
While Fig.~\ref{fig:lz_evolution_withinbar} clearly demonstrates that as the bar grows, there is an associated loss of the angular momentum in the central bar region (for thin and thick discs), it still remains unclear whether the loss in the $l_z$ component display any characteristic azimuthal variation in the radial extent encompassing the bar. To investigate that we compute the two-dimensional distribution of the $l_z$ component in the face-on projection at 5 different times. Fig.~\ref{fig:speclz_2Dmap} shows one such distribution, computed using the thin, thick, and thin+thick disc particles  for the model \textbf{rthickE0.5}. As seen from Fig.~\ref{fig:speclz_2Dmap}, along the bar, there is a prominent preferential deficiency of the specific angular momentum, and this being true for both the thin and thick discs. As the bar grows over time, this preferential deficiency in $l_z$ along the bar becomes more prominent (for both thin and thick discs). Moreover, the thin disc component shows a larger deficiency of the specific angular momentum within the bar region as compared to that for the thick disc component (compare left and middle panels of Fig.~\ref{fig:speclz_2Dmap}). This is not surprising, as the bar in thin disc component is stronger than the bar in the thick disc component, and a stronger bar is expected to transfer angular momentum more vigorously from the disc particles (as also discussed in \citet{Fragkoudietal2017}). We further computed the same two-dimensional distribution of the $l_z$ component in the face-on projection at different times for all other thin+thick disc models considered here. We found that the trend of having a preferential deficit of the specific angular momentum along the bar, remains generic for all thin+thick models.

\subsection{Radial heating of disc particles within the bar region}
\label{sec:radial_heating}

%
\begin{figure*}
\centering
\resizebox{0.86\linewidth}{!}{\includegraphics{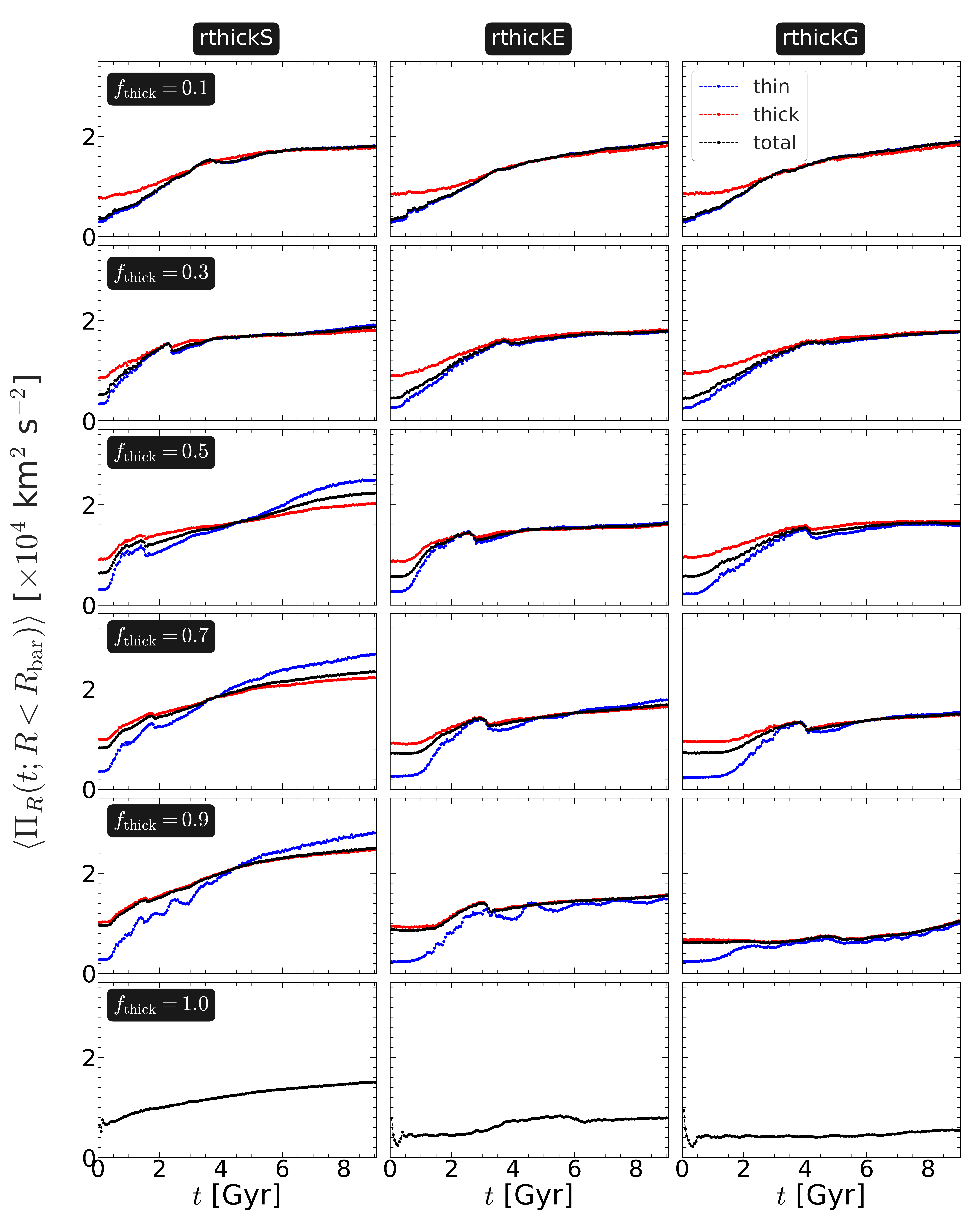}}
\caption{Temporal evolution of the average radial random kinetic energy, calculated within the extent of the bar, $\avg{\Pi_R (t; R < R_{\rm bar})}$ (see Eq.~\ref{eq:avgradialheating_tempEvo}), for the thin (in blue), thick (in red), and total (thin+thick) disc (in black) particles, for all thin+thick disc models. For each model, the $R_{\rm bar}$ was first fixed to a saturation/representative value, for details see text in section.~\ref{sec:angmom_exchange}. \textit{Left-hand panels} show the evolution for the \textbf{rthickS} models whereas  \textit{middle panels} and \textit{right-hand panels} show the evolution for the \textbf{rthickE}  and \textbf{rthickG} models, respectively. The value of $f_{\rm thick}$ varies from 0.1 to 1 (top to bottom), as indicated in the left-most panel of each row. The extents of bars, used in each sub-panel, are already indicated in Fig~\ref{fig:density_maps_endstep_allmodels} (see black solid circles there).}
\label{fig:MeanradialHeating_evolution_withinbar}
\end{figure*}
In the past literature, it is known that a bar heats up the disc materials in both radial and vertical directions \citep[e.g., see][]{Grandetal2016,Pinnaetal2018}. However, here, we investigate how the disc particles in the inner part of the disc gets heated up as the bar grows over time. In the literature, the radial heating is quantified via radial random kinetic energy ($\Pi_R$) which is calculated (within the bar region) using 

\begin{equation}
\Pi_R (t; R < R_{\rm bar}) = \sum_{i=1}^{N(t)} m_i \sigma_R^2\,,
\label{eq:TempEvo_RadialHeating}
\end{equation}
\noindent where $N(t)$ is the total number of stellar particles contained within the bar region at time $t$, and $\sigma_R$ is the radial component of the velocity dispersion. However, again this quantity is a summation over thin or thick disc particles (whichever is applicable), and hence is prone to certain biases  as $f_{\rm thick}$ values vary in our models (see discussion in the above section). To circumvent that, we instead calculate the average radial random kinetic energy ($\avg{\Pi_R}$), within the bar 
\begin{equation}
\avg{\Pi_R (t; R < R_{\rm bar})} = \frac{\sum_{i=1}^{N(t)} m_i \sigma_R^2} {\sum_{i=1}^{N(t)} m_i}\,.
\label{eq:avgradialheating_tempEvo}
\end{equation}
The resulting temporal evolution of $\avg{\Pi_R}$ \footnote{This is basically the specific radial random kinetic energy.} within the bar region, for the thin, thick, and total (thin+thick) disc particles, are shown in Fig.~\ref{fig:MeanradialHeating_evolution_withinbar} for all the thin+thick disc models as well as three thick-disc-only models. As seen in Fig.~\ref{fig:MeanradialHeating_evolution_withinbar}, initially the thin disc remains kinematically colder than the thick disc. However, as the bar grows, the radial mean random kinetic energy for both the disc components (thin and thick) increases monotonically with time. At later times ($t > 6 \Gyr$), when the bar no longer grows appreciably, the corresponding increment in the $\avg{\Pi_R (t; R < R_{\rm bar})}$ also becomes smaller, this being true for both the thin and thick components. These overall trends in the temporal evolution of $\avg{\Pi_R (t; R < R_{\rm bar})}$ hold true for almost all the thin+thick models considered here. Interestingly, by the end of simulation ($t = 9 \Gyr$), the average radial random kinetic energy for the thin and thick particles (within the bar region) becomes similar, although the thin disc was initially kinematically colder than the thick disc. A similar finding was also reported in \citet{Dimatteoetal2019}, the only difference is that in their idealized model, initially the thin and the thick disc share the same vertical velocity dispersion and differ in terms of the in-plane velocity dispersion. For the thick-disc only models, the radial heating of stars, within the bar region, is progressively less from \textbf{rthickS1.0} model to \textbf{rthickE1.0} model to \textbf{rthickG1.0} model (see bottom panels of Fig.~\ref{fig:MeanradialHeating_evolution_withinbar}). This goes with the expectation as the formation of progressively weaker bars should be associated with progressively less radial disc heating in the central bar region.

\par
\begin{figure*}
\centering
\resizebox{0.9\linewidth}{!}{\includegraphics{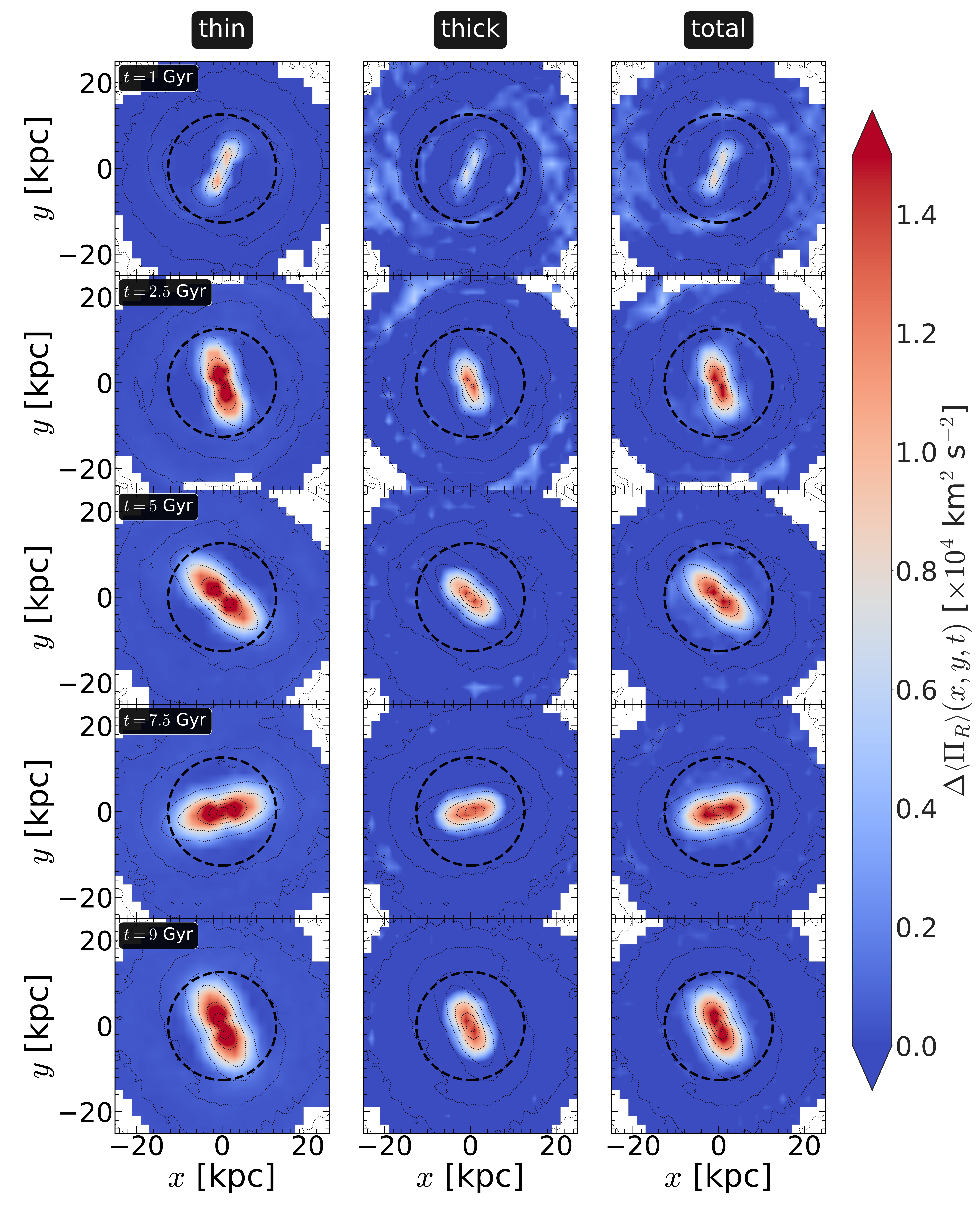}}
\caption{Distribution of the change in the average radial random kinetic energy, $\Delta \left<\Pi_{R} \right > (x, y, t)$ (see Eq.~\ref{eq:Delta_ranKinEnr_2d}), calculated in the face-on projection (($x,y$)-plane), at different times for the thin (left-hand panels), thick (middle panels), and thin+thick disc particles (right-hand panels) for the model \textbf{rthickE0.5}. Dotted lines denote the contours of total (thin+thick) surface density. The black dashed circle in each sub-panel denote the extent of the bar. } 
\label{fig:Difference_Meanrandomheating_2Dmap}
\end{figure*}

Next, we study the two-dimensional distribution of the \textbf{change in the} average random kinetic energy in the face-on projection. In particular, we look for any characteristic variation in the average radial random kinetic energy along the bar. To quantify that, at any time $t$, we define the change in the average radial random kinetic energy as 
\begin{equation}
\Delta \left<\Pi_{R} \right > (x, y, t) =  \left<\Pi_{R} \right > (x, y, t) -  \left<\Pi_{R} \right > (x, y, t=0)\,.
\label{eq:Delta_ranKinEnr_2d}
\end{equation}
In Fig.~\ref{fig:Difference_Meanrandomheating_2Dmap}, the resulting variation of $\Delta \left<\Pi_{R} \right > (x, y, t)$ in the ($x-y$)-plane, calculated for the thin, thick, and thin+thick disc particles \textbf{is} shown for the model \textbf{rthickE0.5}. 
%
As seen clearly from Fig.~\ref{fig:Difference_Meanrandomheating_2Dmap}, there is a preferential excess of radial heating, tracing the spatial two-dimensional extent of the bar (in the face-on projection), this being true for both the thin and the thick disc components. Initially, when the bar is not very strong, the corresponding (preferential) increment of the average radial random kinetic energy is also small. Subsequently, as the bar grows with time, so does the radial random kinetic energy, and by the end of the simulation run ($t = 9 \Gyr$), when the bar is fully grown, the corresponding radial random kinetic energy also reaches its maximum value. Within the bar region, the thin disc particles get more heated up when compared with the thick disc particles. This confirms that developing a stronger bar is associated with a larger degree of radial heating. We further checked this for the other thin+thick models as well. We find that the preferential excess of radial heating along the bar (for both thin and thick components) remains a generic trend.
\par
\begin{figure*}
\centering
\resizebox{\linewidth}{!}{\includegraphics{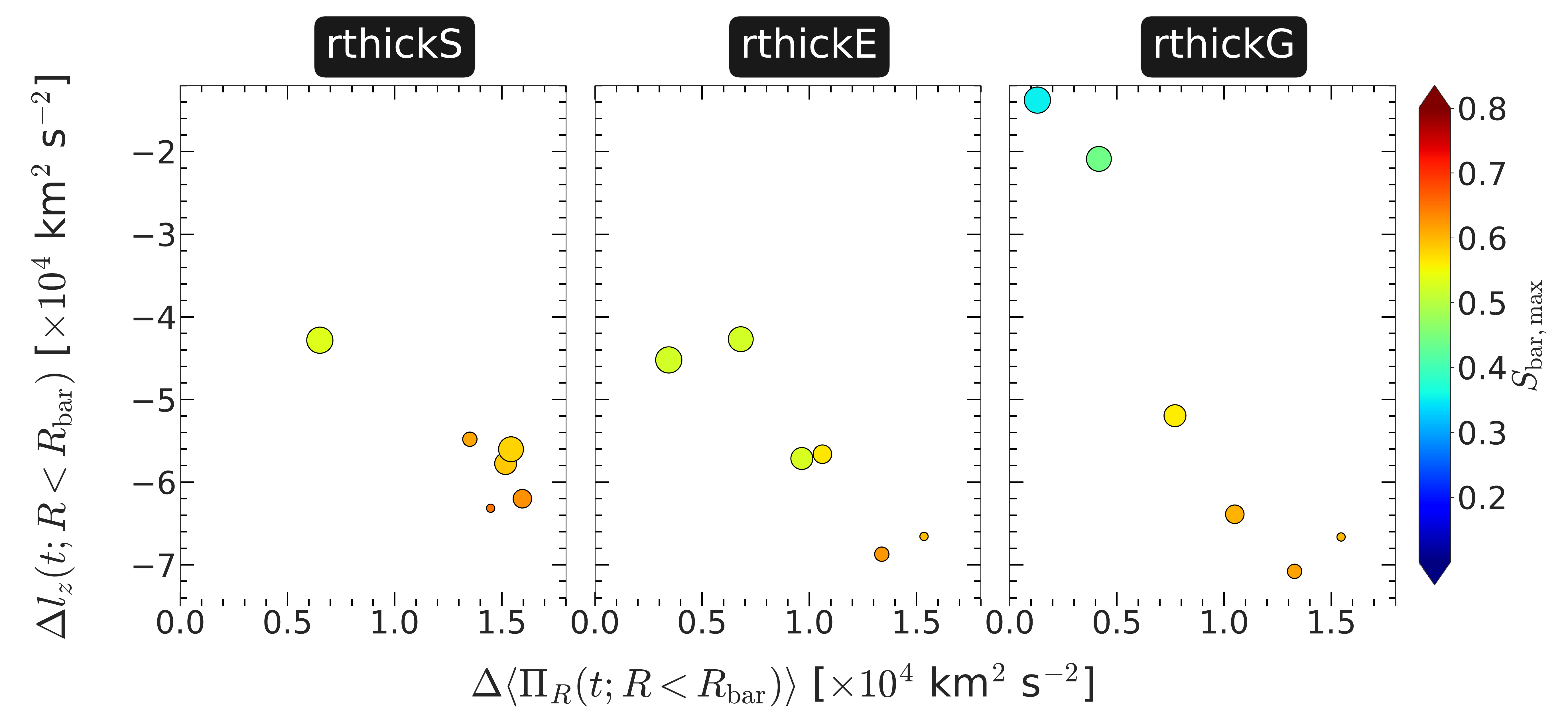}}
\caption{Distribution of change in average radial random kinetic energy and the specific angular momentum, for the total (thin+thick) disc, within the bar region (see Eqs.~\ref{eq:change_in_radialHeating} and \ref{eq:change_in_specAngMom}), for all models considered here. \textit{Left-hand panels} show the evolution for the \textbf{rthickS} models whereas  \textit{middle panels} and \textit{right-hand panels} show the evolution for the \textbf{rthickE}  and \textbf{rthickG} models, respectively. Points are colour-coded by the maximum value of the strength of the bar ($S_{\rm bar, max}$). The increasing size of the points denote higher thick disc fraction ($f_{\rm thick} =0.1-1$).} 
\label{fig:MeanradialHeating_angMomexchange_compare}
\end{figure*}
\par
Lastly, we investigate the correlation (if any) between the changes in the specific angular momentum as well as the radial heating and the bar strength in our considered thin+thick models.
To achieve that, we first compute the change in the specific angular momentum for the total (thin+thick) disc particles, within the bar region, by using
\begin{equation}
\Delta l_z (R< R_{\rm bar})  =  l_z (t_{\rm end}; R< R_{\rm bar}) - l_z(t_0; R< R_{\rm bar})\,,
\label{eq:change_in_specAngMom}
\end{equation}
\noindent and then we also compute the change in the (average) radial heating within the bar region by using
\begin{equation}
\Delta \avg{\Pi_R (R< R_{\rm bar}) } =  \avg{ \Pi_R (t_{\rm end}; R< R_{\rm bar})} - \avg{\Pi_R (t_0; R< R_{\rm bar})}\,.
\label{eq:change_in_radialHeating}
\end{equation}
\noindent This is shown in Fig.~\ref{fig:MeanradialHeating_angMomexchange_compare}. As seen clearly from Fig.~\ref{fig:MeanradialHeating_angMomexchange_compare}, indeed a stronger bar correlates with a higher angular momentum loss and a higher degree of radial heating for stars. This trend holds for a wide variety of configurations (thin-to-thick disc scale length ratio) as well as for different $f_{\rm thick}$ values considered here. This demonstrates that a stronger bar is associated with larger amount of disc heating and larger transfer the angular momentum from the disc particles, in agreement with past findings \citep[e.g. see][]{TremaineWeinberg1984,DebattistaandSellwood2000,Sahaetal2010,Grandetal2016}.

\section{Testing the bar instability criteria}
\label{sec:bar_instability}

Here, we test two bar instability criteria, widely used in the past literature, on all the thin+thick models considered here to explore whether these criteria can indeed reveal when a bar will form.

\subsection {Ostriker-Peebles (OP) criterion}
\label{sec:OP_criterion}

\begin{figure*}
\centering
\resizebox{0.95\linewidth}{!}{\includegraphics{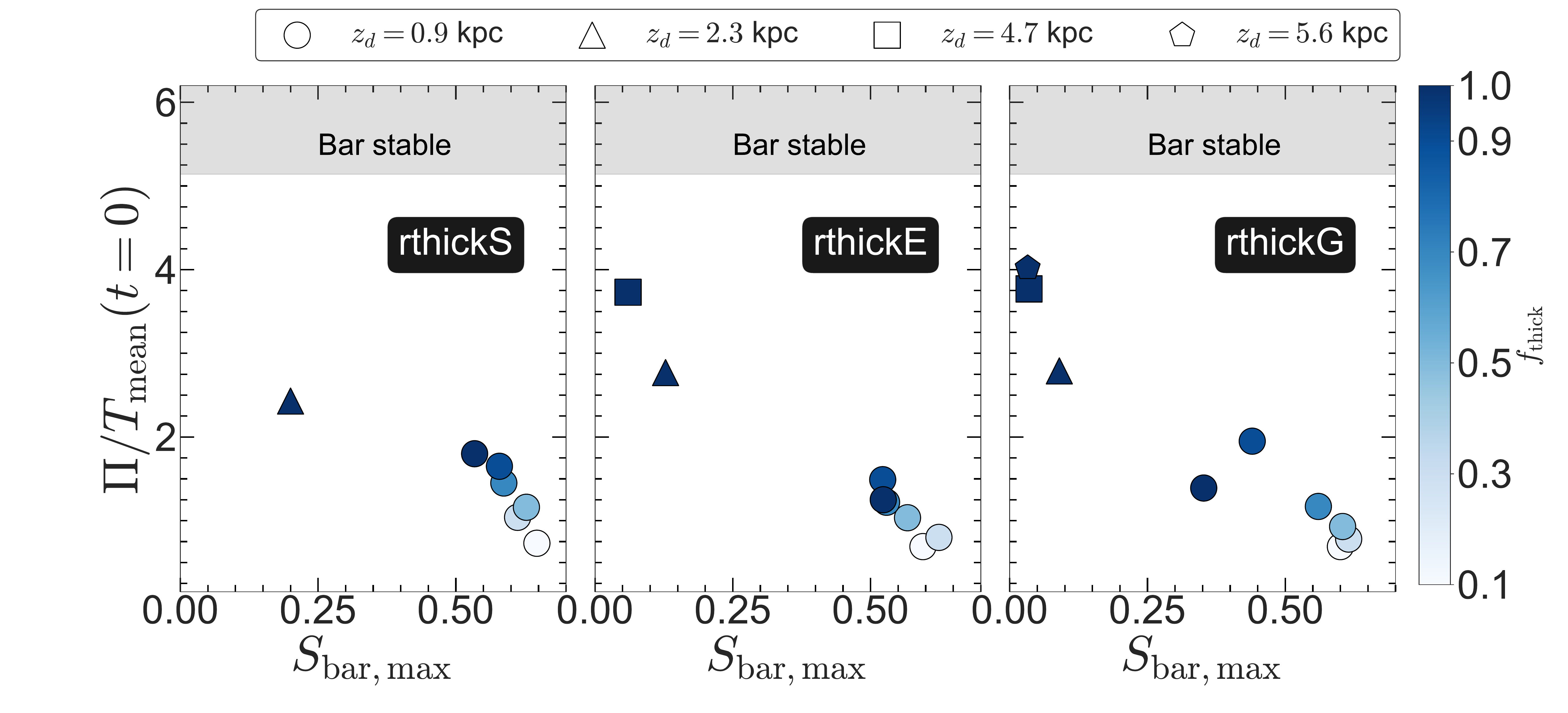}}
\caption{Ostriker-Peeble (OP) criterion : the ratio of the total mean kinetic energy to total random kinetic energy, calculated within the extent of the bar, ($\Pi/T_{\rm mean} (t; R < R_{\rm bar})$) at $t =0$, for the total (thin+thick) disc as a function of maximum bar strength ($S_{\rm bar, max}$). \textit{Left-hand panels} show the evolution for the \textbf{rthickS} models whereas  \textit{middle panels} and \textit{right-hand panels} show the evolution for the \textbf{rthickE}  and \textbf{rthickG} models, respectively. Points are colour-coded by the corresponding thick disc fraction, $f_{\rm thick}$ (see the colour bar). Different symbols represent thin+thick models with different scale heights, for details see text. The shaded region denotes the bar-stable zone, according to the OP criterion.}
\label{fig:op_creterion_allmodels}
\end{figure*}
\begin{figure*}
\centering
\resizebox{0.945\linewidth}{!}{\includegraphics{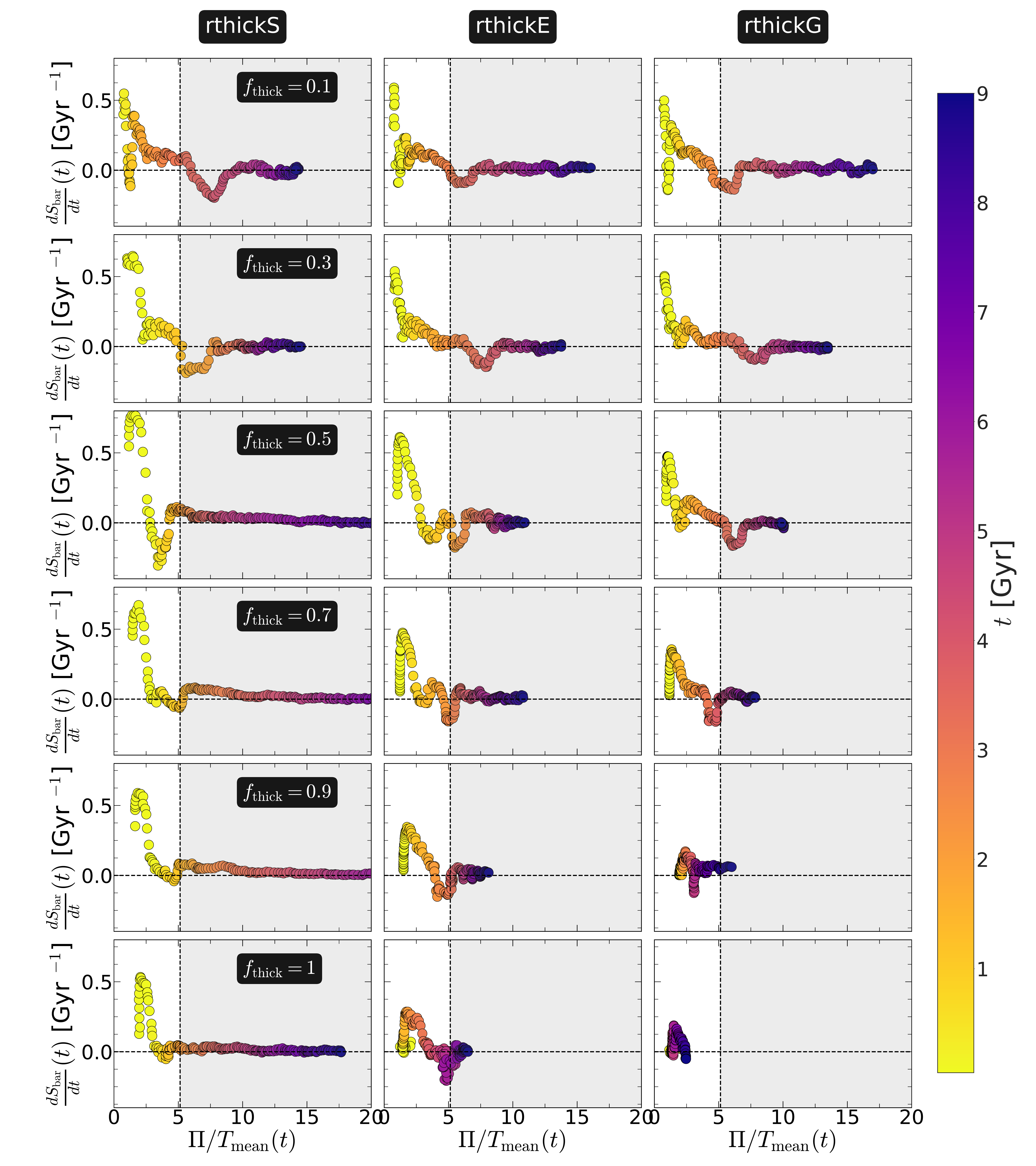}}
\caption{Evolution of the ratio of the total mean kinetic energy to total random kinetic energy, calculated within the extent of the bar, ($\Pi/T_{\rm mean} (t; R < R_{\rm bar})$), for the total (thin+thick) disc particles, plotted against the growth rate of the bar ($d S_{\rm bar}/dt $) for all thin+thick models with varying $f_{\rm thick}$ values. \textit{Left-hand panels} show the evolution for the \textbf{rthickS} models whereas  \textit{middle panels} and \textit{right-hand panels} show the evolution for the \textbf{rthickE}  and \textbf{rthickG} models, respectively. The thick disc fraction ($f_{\rm thick}$) varies from 0.1 to 1 (top to bottom), as indicated in the left-most panel of each row. The points are colour-coded by the simulation time (see the colour bar). The vertical black line denotes $\Pi/T_{\rm mean} = 5.14$ which serves as a boundary for the bar instability phase, and the grey shaded region (in each sub-panel) denotes the bar-stable phase according to the OP criterion, for details see text.}
\label{fig:OPcriterion_evolution}
\end{figure*}
Using $N$-body simulations of disc galaxies, the seminal work by \citet{OstrikerandPeebles1973} showed that a stellar disc would enter into the bar instability phase if the ratio of the total mean kinetic energy to the potential energy, $W$, exceeds a critical limit of $0.14 \pm 0.003$. Assuming the Virial theorem holds for our simulation snapshots, this criterion can well be converted in terms of the rotational kinetic energy and the random kinetic energy.
The random kinetic energy is calculated using 
\begin{equation}
\Pi(R) = \Pi_R(R) + \Pi_{\phi}(R)+\Pi_z(R)\,,
\end{equation}
\noindent where each component of the random kinetic energy is calculated as \citep{BinneyTremaine2008}
\begin{equation}
\Pi_j(R) = \sum_{i=1}^{N_R} m(i) \sigma_j^2(R)\,.
\end{equation}
\noindent Here, $j = R, \phi, z$; $\sigma_j$ is the corresponding velocity dispersion, and $m(i)$ is the mass of the $i$th disc particle. Similarly, the total mean kinetic energy is calculated as 
\begin{equation}
T_{\rm mean}(R) = T_R(R) + T_{\phi}(R)+T_z(R)\,,
\end{equation}
\noindent where each component of the mean kinetic energy is calculated as \citep{BinneyTremaine2008}
\begin{equation}
T_j(R) = \sum_{i=1}^{N_R} \frac{1}{2}m(i) \avg{v_j(R)}^2\,.
\end{equation}
\noindent Here also, $j = R, \phi, z$; $\avg{v_j}$ is the corresponding mean velocity, and $N_R$ is the total number of particles in the radial extent $[R, R+\Delta R]$. Now, following the tensor-Virial theorem in a steady state, taking the trace of the kinetic and the potential tensors we get \citep{BinneyTremaine2008}
\begin{equation}
\Pi + 2 T_{\rm mean}+ W = 0\,.
\end{equation} 
\noindent Then, applying the OP criterion, we get \citep[for details, see][]{SahaElmegreen2018}
\begin{equation}
\frac{T_{\rm mean}}{|\Pi + 2T_{\rm mean}|} > 0.14\,,
\end{equation}
\noindent or equivalently,
\begin{equation}
\Pi/T_{\rm mean} +2 < 7.14 =>  \Pi/T_{\rm mean} < 5.14\,.
\label{eq:Ostriker_peebleCriterion}
\end{equation}
\noindent To explain, if the ratio $\Pi/T_{\rm mean}$ is less than $5.14$, the stellar disc undergoes the bar instability phase, and if the ratio $\Pi/T_{\rm mean}$ is greater than $5.14$, the corresponding stellar disc remains stable against the bar formation. 
%
\par
In Fig.~\ref{fig:op_creterion_allmodels}, we show the ratio $\Pi/T_{\rm mean}$, calculated at $t =0$, within the extent of the bar, plotted against the maximum values of the bar strength for all 25 models considered here.  The $\Pi/T_{\rm mean} (t=0)$ values for the 15 thin+thick models which eventually develop a strong bar, lie below $5.14$, thereby are qualified for being bar unstable according to the OP criterion. The OP criterion also successfully predicts the thick-disc-only models, namely \textbf{rthickS1.0}, \textbf{rthickE1.0}, and \textbf{rthickG1.0} as bar-unstable (with $\Pi/T_{\rm mean} (t =0) < 5.14$). However, the OP criterion fails to predict correctly the bar stability scenario for the thick-disc-only models where we have increased the disc scale height. In other words, the $\Pi/T_{\rm mean} (t =0) $ values for these models remain less than $5.14$, thereby qualified as being bar-unstable according to the the OP criterion. However, most of these thick-disc-only models (with larger scale height) do not form a prominent bar at all within the simulation run ($9 \Gyr$). Nevertheless, we found an anti-correlation between the $\Pi/T_{\rm mean} (t =0) $ values and the maximum bar strength ($S_{\rm bar, max}$) which is expected in the sense that a system with lower $\Pi/T_{\rm mean} (t =0) $ indeed develops a stronger bar in the disc (see Fig.~\ref{fig:op_creterion_allmodels}). We further checked if there is any correlation between the temporal evolution of $\Pi/T_{\rm mean} (t)$ and the growth rate of the bar. This \textbf{is} shown in Fig.~\ref{fig:OPcriterion_evolution} for the models considered here. As seen clearly, for most of the models, the growth rate of the bar stabilises (i.e., $d S_{\rm bar}/dt \sim 0$) after $\Pi/T_{\rm mean} (t)$ crosses the value 5.14. In other words, when the system eventually becomes bar stable, according to the OP criterion, the bar does not grow drastically either (although it might go through the buckling phase). While this trend remains true for most of the models, we find  $\Pi/T_{\rm mean} (t)$ stays low for the models that do not form a strong bar, e.g., \textbf{rthickG1.0} (bottom right panel in Fig.~\ref{fig:OPcriterion_evolution}). One plausible explanation could be since these models do not form a strong bar, therefore, the disc stars do not get heated up by the action of a (strong) bar. Hence, the total random kinetic energy remains low, which, in turn, keeps the ratio $\Pi/T_{\rm mean}$ smaller or closer to 5.14. In addition, for some models (see bottom left panels of Fig.~\ref{fig:OPcriterion_evolution}), the bar strength does not increase even though the values of $\Pi/T_{\rm mean}$ increase steadily.

\subsection {Efstathiou-Lake-Negroponte (ELN)  Criterion}
\label{sec:efstathiou_criterion}
\begin{figure*}
\centering
\resizebox{0.95\linewidth}{!}{\includegraphics{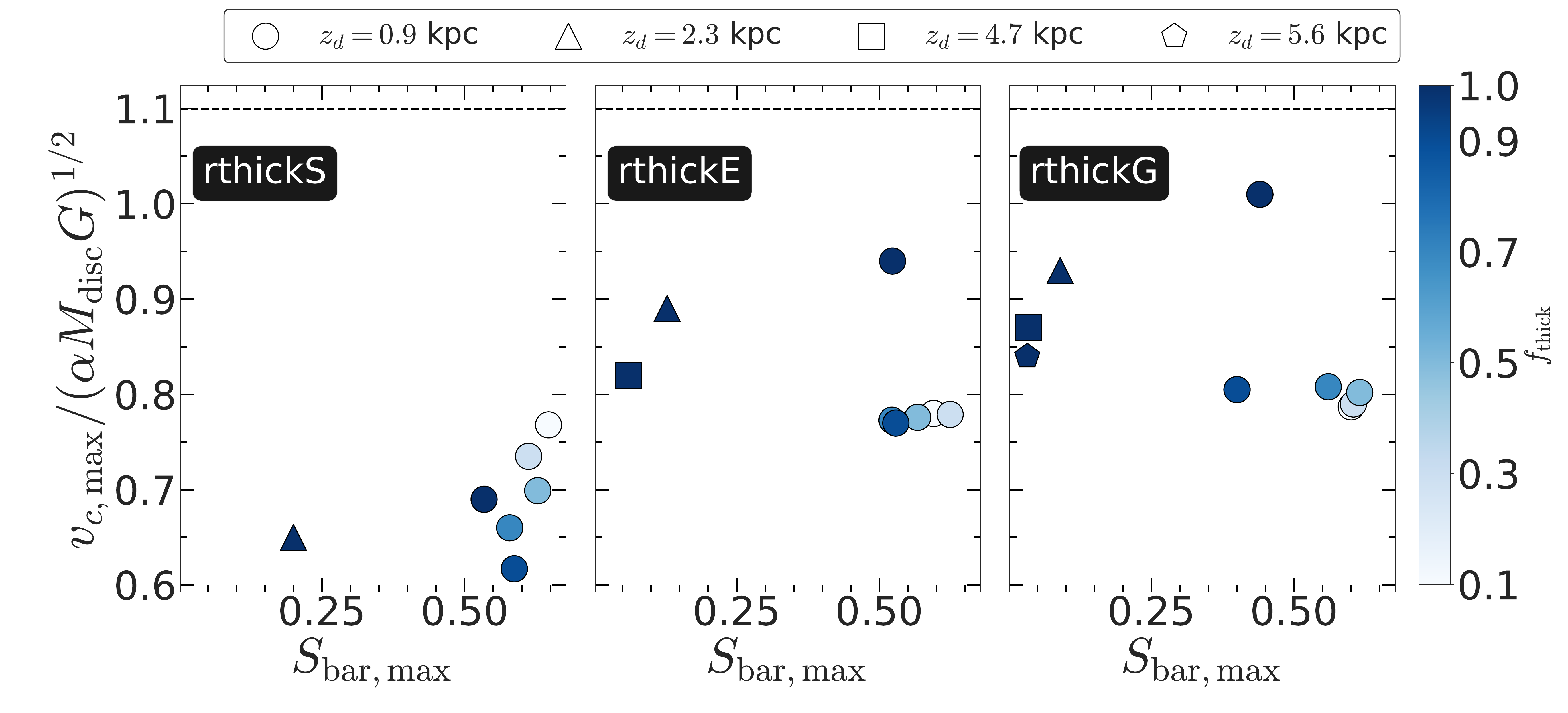}}
\caption{Efstathiou-Lake-Negroponte (ELN) criterion : the quantity, $v_{c, \rm max}/(\alpha M_{\rm disc} G)^{1/2}$, calculated at $t=0$, for all thin+thick disc models, as a function of maximum bar strength ($S_{\rm bar, max}$). \textit{Left-hand panels} show the evolution for the \textbf{rthickS} models whereas  \textit{middle panels} and \textit{right-hand panels} show the evolution for the \textbf{rthickE}  and \textbf{rthickG} models, respectively. Points are colour-coded by the corresponding thick disc fraction, $f_{\rm thick}$ (see the colour bar). Different symbols represent thin+thick models with different scale heights. The horizontal line (in black) denotes $v_{c, \rm max}/(\alpha M_{\rm disc} G)^{1/2} = 1.1$ which serves as a boundary for the bar instability phase, for details see text. }
\label{fig:efstathiou_creterion_allmodels}
\end{figure*}

\citet{Efstathiouetal1982} proposed an analytic criterion to determine the stability of a stellar disc against the bar formation :
\begin{equation}
\varepsilon = \frac{v_{\rm c, max}}{(\alpha M_{\rm disc} G)^{1/2}}\,,
\end{equation}
\noindent where $v_{\rm c, max}$ is the maximum circular velocity, $M_{\rm disc}$ is the total disc mass, and $\alpha = R^{-1}_{\rm d}$, $R_{\rm d}$ being the scale-length of the stellar disc. A stellar disc would become bar unstable for $\varepsilon \leq 1.1$, and if $\varepsilon > 1.1$, the stellar disc would be stable against the bar formation \citep[for details see][]{Efstathiouetal1982}.
\par
To compute $\varepsilon$ at $t=0$ for the total (thin+thick) disc, first we calculated the initial ($t =0$) circular velocity profiles for the whole system, and then determined the peak value of the circular velocity. The radial profiles circular velocity (or equivalently, the rotation curve), at $t=0$, is first derived from the underlying (equilibrium) mass distribution at the initial time. Then, from the radial variation, we take the maximum value of the $v_{\rm c}$ as $v_{\rm c, max}$. As for the determination of $R_{\rm d}$, we point out that our thin+thick models consist of a thin and a thick disc, with varying disc scale lengths (see Table~\ref{table:key_param}). Therefore, we compute the average disc scale length, $\avg{R_{\rm d}}$, in the following fashion
\begin{equation}
\avg{R_{\rm d}}  = \frac{M_{\rm d, thin} R_{\rm d, thin} + M_{\rm d, thick} R_{\rm d, thick}} {M_{\rm d, thin}+M_{\rm d, thick}}\,.
\label{eq:avg_scalelength}
\end{equation}

Fig~\ref{fig:efstathiou_creterion_allmodels} shows the resulting values of the dimensionless quantity, $\varepsilon$, corresponding to the equilibrium models ($t =0$), plotted against the maximum values of the bar strength for all 25 models considered here. The $\varepsilon$ values for the 15 thin+thick models which eventually develop a strong bar, lie below $1.1$, thereby are qualified for being bar unstable according to the ELN criterion. The ELN criterion also successfully predicts the thick-disc-only models, namely \textbf{rthickS1.0}, \textbf{rthickE1.0}, and \textbf{rthickG1.0} as bar-unstable (with $\varepsilon <1.1$). However, the ELN criterion fails to predict correctly the bar stability scenario for the thick-disc-only models where we have increased the disc scale height. In other words, the $\varepsilon$ values for these models remain less than $1.1$, thereby qualified as being bar-unstable according to the the ELN criterion. However, most of these thick-disc-only models (with larger scale height) do not form a prominent bar at all within the simulation run ($9 \Gyr$). Moreover, we did not find any anti-correlation between the $\varepsilon$ values and the maximum bar strength ($S_{\rm bar, max}$) which was expected in the sense that a system with lower $\varepsilon$ value should have developed a stronger bar in the disc.

\section{Discussion}
\label{sec:discussion}
\subsection{Caveat to our analysis}
%
In what follows we discuss some of the implications and limitations of this work. First, in the thin+thick models considered here, the stars are separated into two well-defined and distinct populations, namely, the thin-disc stars and the thick-disc stars. We point out that this scheme of segregating stars into two distinct populations might be suitable for the external galaxies which have two discrete disc structures, however this is a simplification for the Milky Way, where stars of different mono-abundance populations show a smooth transition from a thin to a thick disc. \citet{Bovyetal2012} showed that disc properties vary continuously with the scale height. Therefore, treating the stars, separated into two distinct populations, reduces the complexity that is expected to be present in the Milky Way. Nevertheless, this discretised treatment of stars with varying fraction of thick disc stars provides valuable insight regarding the trends that will be followed by the stellar populations in the external disc galaxies. We further mention that some of the findings of this work can be extended to more complex configurations. For example, the similar radial velocity dispersion for the thin and the thick disc, at the end of the simulation run, as shown here, was also found in \citet{Dimatteoetal2019} where the galaxy was modelled by three components (thin, intermediate, and thick discs).
\par
The radial extents of the thin and thick discs merit some discussion. The extent of the thick disc partially depend on how it is defined. To illustrate, for the Milky Way, when the thick disc component is defined in terms of the $\alpha$-enhanced stars, the scale length of thick disc is smaller than that for the thin disc. However, when the classification for the thick disc is based on colours, and not the chemical abundances, the thick disc scale length is larger than that of the thin disc \citep[e.g., see][]{Juricetal2008}. Also in external galaxies, the thick discs are generally more extended than thin disc \citep[e.g., see][]{Yoachim2006}. This motivates us to consider all three possibilities regarding the thin-to-thick disc scale length ratio (\textbf{rthickS}, \textbf{rthickE}, and \textbf{rthickG} models), and address the bar formation scenario for a wide range of thin-to-thick disc geometric configurations.
\par
Furthermore, the thin+thick models used here, are collisionless simulations, and do not contain any interstellar gas. The dynamical role of the interstellar gas in the context of generation/destruction of disc instabilities, e.g., bars \citep{Bournaudetal2005} and spiral arms \citep{SellwoodCarlberg1984,GhoshJog2015,GhoshJog2016,GhoshJog2021} has been investigated in the past literature. The presence of a (dynamically) cold component, e.g., gas, makes the disc more susceptible to the gravitational instabilities \citep[e.g., see][]{JogandSolomon1984,Jog1996,Bertin2000}. Furthermore, \citet{Bournaudetal2005} argued that the inflow of the gas in the central region infuses angular momentum, thereby leading to significant bar weakening phenomenon. However, recent observational work by \citet{Pahwa2018} showed the presence of prominent bars in several low-surface-brightness (LSB) galaxies with high gas fraction. Nevertheless, it will be worth investigating the bar formation scenario in presence of both the thick disc and the interstellar gas. 
\par
\subsection{On the formation of bars in thick discs}
%
As for the formation of bars, the simulations analysed in this paper show that a bar can form also in thick-disc dominated galaxies or in pure thick-disc galaxies, for sufficiently compact distributions (in terms of scale lengths and heights). In particular,  our models show that a bar could have formed also in thick discs with characteristics  similar to those of the $\alpha$-enhanced thick disc of the Milky Way \citep[scale length of $2 \kpc$, scale height of $0.9 \kpc$; see][]{Bovyetal2012}, even if the thin disc counterpart was not in place yet.  For a galaxy like the Milky Way, the formation of the thin disc started about $9 \Gyr$ ago \citep[$z \sim 1$, see][]{Haywoodetal2013,Snaithetal2014}. This suggests that the Milky Way's bar may have been forming or already in place at those epochs. This finding may, in principle, push to earlier epochs than those suggested so far \citep{Haywoodetal2016,Bovyetal2019} the time of the bar formation in our Galaxy, with an upper limit to its formation possibly set by the end of the epoch of significant mergers experienced by the Milky Way, currently estimated at about $10 \pm 1 \Gyr$ ago \citep[see][]{Belokurovetal2018,Dimatteoetal2019,Kruijssenetal2020}. As discussed previously, it would be crucial to investigate how the inclusion of gas in the models can modify these suggestions. Furthermore, recent observations of Rest-Frame Near-Infrared images with the JWST showed the clear presence of prominent bars in disc galaxies in the high redshift universe ($z > 1$) as well \citep{Guoetal2022}. In such redshift ranges, the disc galaxies are believed to be clumpy, turbulent, and more importantly, having thick discs \citep[e.g., see][]{Hamilton-Campos+2023}. The results presented here, clearly demonstrated that bar formation in presence of thick discs is possible, even when the thick disc stars dominate the underlying mass distribution. Therefore, our study here provides a natural explanation for the bar formation scenario in high redshift galaxies, as recently unveiled by the JWST observations.

\subsection{Criteria for bar formation}
As shown earlier, the OP criterion fails to predict the bar formation scenario correctly for some cases, especially for those thick-disc-only models with larger scale height. However, this is not surprising as the OP criterion does not include all the complexity related to bar formation and its growth. For example, the formalism does not take into account the interaction between the disc and the dark matter halo. The limiting value of ratio of the rotational kinetic energy to the potential energy, $W$, exceeding $0.14 \pm 0.003$ in order to become bar-stable were derived from simulations which had rigid dark matter halo, and not the live dark matter halo \citep[also see discussion in][]{Athanassoula2008}.
\par
The limited success of the predictability of the ELN criterion for our models is also not surprising, as the ELN criterion was originally developed based on 2-D simulation models, and hence it does not take into account the (destabilising) effect of the disc-dark matter halo interaction. Furthermore, it does not take into account the (stabilising) effect of the disc velocity dispersion or the central concentration of the dark matter halo \citep[for details, see discussions in][]{Athanassoula2008,Romeoetal2022}. Recent study of \citet{Romeoetal2022} showed that for their sample of selected  observed galaxies, in only 50-55 percent cases, the ELN criterion successfully predicts the bar formation scenario. On the other hand, the study of \citet{Izquierdo-Villalbaetal2022} showed that for around 70-80 percent of the galaxies, taken from the IllustrisTNG simulations, the ELN criterion correctly predicts the bar formation scenario.

\section{Summary}
\label{sec:conclusion}

In summary, we investigated the dynamical effect of a geometrically thick disc on the bar formation and evolution scenario. We constructed a suite of $N$-body models of thin+thick discs while systematically varying mass fraction of thick disc as well as for different thin-to-thick disc scale length ratios. Our main findings are :\\

\begin{itemize}

\item{Bars form in almost all thin+thick disc models with varying thick disc mass fraction and for all three geometric configurations with different thin-to-thick disc scale length ratios. The bar is present in both thin and thick disc stars. The bar in the thick disc always remains weaker than the bar in the thin disc stars. Nevertheless, the overall trend of bar evolution and growth in both thin and thick discs mimics each other.}

\item{Bars form quite early ($\sim 2 \Gyr$) in almost all thin+thick disc models considered here. The models with thick disc scale length shorter than that for the thin disc (\textbf{rthickS} models) form the bar relatively earlier when compared to the other two geometric configurations. In addition, the bars in \textbf{rthickS} models display a more rapid growth in the initial phase as compared to other two geometric configurations.}

\item{The bar formation in thick-disc-only simulations (without any thin disc) depends critically on the scale length and scale height of the thick disc. With increasing scale length (and for a fixed scale height), the bar formation epoch gets delayed, and the resulting bar is progressively weaker. Similarly, with increasing scale height (and for a fixed scale length), the system becomes progressively hostile to bar formation.}

\item{Formation of a stronger bar is associated with a larger angular momentum loss and the radial heating within the bar region, in agreement with previous findings. Moreover, we demonstrate a preferential loss of the angular momentum and a preferentially increased radial heating, along the two-dimensional extent of the bar. This trend holds true for both the thin and thick disc stars, and remains generic for all thin+thick models considered here.}

\item{We find the Ostriker-Peeble (OP) criterion is a better prescription than the Efstathiou-Lake-Negroponte (ELN) criterion for predicting bar formation scenarios in our thin+thick models. While the Ostriker-Peeble criterion almost always predicts correctly the bar instability in our models, the success of predicting bar instability from the Efstathiou-Lake-Negroponte criterion remains only limited for our models.}

\end{itemize}

To conclude, even the inclusion of a massive (kinematically hot) thick disc component is not efficient in suppressing the bar instability, suggesting that bars can form in hot thick discs at high redshifts. These results are in agreement with the recent observational evidence from the JWST, and provides a natural explanation for the bar formation scenario in disc galaxies at the high redshift universe ($z > 1$). Our study demonstrates that if the disc galaxy consists only of a thick disc, bar formation can be delayed/ prevented (within a Hubble time) only if the disc scale length is larger (i.e., with less centrally-concentrated surface density) and/or if the vertical scale height is larger (i.e., more vertically extended and kinematically hotter) in disc galaxies.

\section*{Acknowledgements}
We thank the anonymous referee for useful comments which helped to improve this paper. S.G. acknowledges funding from the Alexander von Humboldt Foundation, through Dr. Gregory M. Green's Sofja Kovalevskaja Award.This work has made use of the computational resources obtained through the DARI grant A0120410154 (P.I. : P. Di Matteo).

\bibliographystyle{aa.bst} 
\bibliography{my_ref.bib} 

\begin{thebibliography}{108}
\expandafter\ifx\csname natexlab\endcsname\relax\def\natexlab#1{#1}\fi

\bibitem[{{Aguerri} {et~al.}(2009){Aguerri}, {M{\'e}ndez-Abreu}, \&
  {Corsini}}]{Aguerrietal2009}
{Aguerri}, J.~A.~L., {M{\'e}ndez-Abreu}, J., \& {Corsini}, E.~M. 2009, \aap,
  495, 491

\bibitem[{{Aguerri} {et~al.}(2000){Aguerri}, {Mu{\~n}oz-Tu{\~n}{\'o}n},
  {Varela}, \& {Prieto}}]{Aguerrietal2000}
{Aguerri}, J.~A.~L., {Mu{\~n}oz-Tu{\~n}{\'o}n}, C., {Varela}, A.~M., \&
  {Prieto}, M. 2000, \aap, 361, 841

\bibitem[{{Athanassoula}(2003)}]{Athanassoula2003}
{Athanassoula}, E. 2003, \mnras, 341, 1179

\bibitem[{{Athanassoula}(2008)}]{Athanassoula2008}
{Athanassoula}, E. 2008, \mnras, 390, L69

\bibitem[{{Athanassoula} {et~al.}(2005){Athanassoula}, {Lambert}, \&
  {Dehnen}}]{Athanassoulaetal2005}
{Athanassoula}, E., {Lambert}, J.~C., \& {Dehnen}, W. 2005, \mnras, 363, 496

\bibitem[{{Athanassoula} {et~al.}(2013){Athanassoula}, {Machado}, \&
  {Rodionov}}]{Athanassoula2013}
{Athanassoula}, E., {Machado}, R. E.~G., \& {Rodionov}, S.~A. 2013, \mnras,
  429, 1949

\bibitem[{{Athanassoula} \& {Misiriotis}(2002)}]{Athanassoula2002}
{Athanassoula}, E. \& {Misiriotis}, A. 2002, \mnras, 330, 35

\bibitem[{{Athanassoula} \& {Sellwood}(1986)}]{AthanassoulaandSellwood1986}
{Athanassoula}, E. \& {Sellwood}, J.~A. 1986, \mnras, 221, 213

\bibitem[{Aumer \& Binney(2017)}]{AumerandBinney2017}
Aumer, M. \& Binney, J. 2017, Monthly Notices of the Royal Astronomical
  Society, 470, 2113

\bibitem[{{Barnes} \& {Hut}(1986)}]{BarnesandHut1986}
{Barnes}, J. \& {Hut}, P. 1986, \nat, 324, 446

\bibitem[{{Barway} {et~al.}(2011){Barway}, {Wadadekar}, \&
  {Kembhavi}}]{Barwayetal2011}
{Barway}, S., {Wadadekar}, Y., \& {Kembhavi}, A.~K. 2011, \mnras, 410, L18

\bibitem[{{Belokurov} {et~al.}(2018){Belokurov}, {Erkal}, {Evans}, {Koposov},
  \& {Deason}}]{Belokurovetal2018}
{Belokurov}, V., {Erkal}, D., {Evans}, N.~W., {Koposov}, S.~E., \& {Deason},
  A.~J. 2018, \mnras, 478, 611

\bibitem[{{Bertin}(2000)}]{Bertin2000}
{Bertin}, G. 2000, {Dynamics of Galaxies}

\bibitem[{{Binney} \& {Tremaine}(2008)}]{BinneyTremaine2008}
{Binney}, J. \& {Tremaine}, S. 2008, {Galactic Dynamics: Second Edition}
  (Princeton University Press)

\bibitem[{{Bournaud} {et~al.}(2005){Bournaud}, {Combes}, \&
  {Semelin}}]{Bournaudetal2005}
{Bournaud}, F., {Combes}, F., \& {Semelin}, B. 2005, \mnras, 364, L18

\bibitem[{{Bovy} {et~al.}(2019){Bovy}, {Leung}, {Hunt}, {Mackereth},
  {Garc{\'\i}a-Hern{\'a}ndez}, \& {Roman-Lopes}}]{Bovyetal2019}
{Bovy}, J., {Leung}, H.~W., {Hunt}, J. A.~S., {et~al.} 2019, \mnras, 490, 4740

\bibitem[{{Bovy} {et~al.}(2012){Bovy}, {Rix}, {Liu}, {Hogg}, {Beers}, \&
  {Lee}}]{Bovyetal2012}
{Bovy}, J., {Rix}, H.-W., {Liu}, C., {et~al.} 2012, \apj, 753, 148

\bibitem[{{Bovy} {et~al.}(2016){Bovy}, {Rix}, {Schlafly}, {Nidever},
  {Holtzman}, {Shetrone}, \& {Beers}}]{Bovyetal2016}
{Bovy}, J., {Rix}, H.-W., {Schlafly}, E.~F., {et~al.} 2016, \apj, 823, 30

\bibitem[{{Burstein}(1979)}]{Burstein1979}
{Burstein}, D. 1979, \apj, 234, 829

\bibitem[{{Buta} {et~al.}(2010){Buta}, {Laurikainen}, {Salo}, \&
  {Knapen}}]{Butaetal2010}
{Buta}, R., {Laurikainen}, E., {Salo}, H., \& {Knapen}, J.~H. 2010, \apj, 721,
  259

\bibitem[{{Combes} {et~al.}(1990){Combes}, {Debbasch}, {Friedli}, \&
  {Pfenniger}}]{Combesetal1990}
{Combes}, F., {Debbasch}, F., {Friedli}, D., \& {Pfenniger}, D. 1990, \aap,
  233, 82

\bibitem[{{Comer{\'o}n} {et~al.}(2011{\natexlab{a}}){Comer{\'o}n}, {Elmegreen},
  {Knapen}, {Salo}, {Laurikainen}, {Laine}, {Athanassoula}, {Bosma}, {Sheth},
  {Regan}, {Hinz}, {Gil de Paz}, {Men{\'e}ndez-Delmestre}, {Mizusawa},
  {Mu{\~n}oz-Mateos}, {Seibert}, {Kim}, {Elmegreen}, {Gadotti}, {Ho},
  {Holwerda}, {Lappalainen}, {Schinnerer}, \& {Skibba}}]{Comenronetal2011a}
{Comer{\'o}n}, S., {Elmegreen}, B.~G., {Knapen}, J.~H., {et~al.}
  2011{\natexlab{a}}, \apj, 741, 28

\bibitem[{{Comer{\'o}n} {et~al.}(2011{\natexlab{b}}){Comer{\'o}n}, {Knapen},
  {Sheth}, {Regan}, {Hinz}, {Gil de Paz}, {Men{\'e}ndez-Delmestre},
  {Mu{\~n}oz-Mateos}, {Seibert}, {Kim}, {Athanassoula}, {Bosma}, {Buta},
  {Elmegreen}, {Ho}, {Holwerda}, {Laurikainen}, {Salo}, \&
  {Schinnerer}}]{Comeron2011b}
{Comer{\'o}n}, S., {Knapen}, J.~H., {Sheth}, K., {et~al.} 2011{\natexlab{b}},
  \apj, 729, 18

\bibitem[{{Comer{\'o}n} {et~al.}(2018){Comer{\'o}n}, {Salo}, \&
  {Knapen}}]{Comeronetal2018}
{Comer{\'o}n}, S., {Salo}, H., \& {Knapen}, J.~H. 2018, \aap, 610, A5

\bibitem[{{Comer{\'o}n} {et~al.}(2019){Comer{\'o}n}, {Salo}, {Knapen}, \&
  {Peletier}}]{Comeronetal2019}
{Comer{\'o}n}, S., {Salo}, H., {Knapen}, J.~H., \& {Peletier}, R.~F. 2019,
  \aap, 623, A89

\bibitem[{{Comer{\'o}n} {et~al.}(2016){Comer{\'o}n}, {Salo}, {Peletier}, \&
  {Mentz}}]{Comeronetal2016}
{Comer{\'o}n}, S., {Salo}, H., {Peletier}, R.~F., \& {Mentz}, J. 2016, \aap,
  593, L6

\bibitem[{{Contopoulos} \& {Grosbol}(1989)}]{ContopoulosandGrosbol1989}
{Contopoulos}, G. \& {Grosbol}, P. 1989, \aapr, 1, 261

\bibitem[{{Debattista} {et~al.}(2017){Debattista}, {Ness}, {Gonzalez},
  {Freeman}, {Zoccali}, \& {Minniti}}]{Debattistaetal2017}
{Debattista}, V.~P., {Ness}, M., {Gonzalez}, O.~A., {et~al.} 2017, \mnras, 469,
  1587

\bibitem[{{Debattista} \& {Sellwood}(1998)}]{DebattistaandSellwood1998}
{Debattista}, V.~P. \& {Sellwood}, J.~A. 1998, \apjl, 493, L5

\bibitem[{{Debattista} \& {Sellwood}(2000)}]{DebattistaandSellwood2000}
{Debattista}, V.~P. \& {Sellwood}, J.~A. 2000, \apj, 543, 704

\bibitem[{{Di Matteo} {et~al.}(2019){Di Matteo}, {Fragkoudi}, {Khoperskov},
  {Ciambur}, {Haywood}, {Combes}, \& {G{\'o}mez}}]{Dimatteoetal2019}
{Di Matteo}, P., {Fragkoudi}, F., {Khoperskov}, S., {et~al.} 2019, \aap, 628,
  A11

\bibitem[{{Dubinski} {et~al.}(2009){Dubinski}, {Berentzen}, \&
  {Shlosman}}]{Dubinskietal2009}
{Dubinski}, J., {Berentzen}, I., \& {Shlosman}, I. 2009, \apj, 697, 293

\bibitem[{{Efstathiou} {et~al.}(1982){Efstathiou}, {Lake}, \&
  {Negroponte}}]{Efstathiouetal1982}
{Efstathiou}, G., {Lake}, G., \& {Negroponte}, J. 1982, \mnras, 199, 1069

\bibitem[{{Elmegreen} {et~al.}(2004){Elmegreen}, {Elmegreen}, \&
  {Hirst}}]{Elmetal2004}
{Elmegreen}, B.~G., {Elmegreen}, D.~M., \& {Hirst}, A.~C. 2004, \apj, 612, 191

\bibitem[{{Eskridge} {et~al.}(2000){Eskridge}, {Frogel}, {Pogge}, {Quillen},
  {Davies}, {DePoy}, {Houdashelt}, {Kuchinski}, {Ram{\'\i}rez}, {Sellgren},
  {Terndrup}, \& {Tiede}}]{Eskridgeetal2000}
{Eskridge}, P.~B., {Frogel}, J.~A., {Pogge}, R.~W., {et~al.} 2000, \aj, 119,
  536

\bibitem[{{Fragkoudi} {et~al.}(2017){Fragkoudi}, {Di Matteo}, {Haywood},
  {G{\'o}mez}, {Combes}, {Katz}, \& {Semelin}}]{Fragkoudietal2017}
{Fragkoudi}, F., {Di Matteo}, P., {Haywood}, M., {et~al.} 2017, \aap, 606, A47

\bibitem[{{Fragkoudi} {et~al.}(2020){Fragkoudi}, {Grand}, {Pakmor},
  {Bl{\'a}zquez-Calero}, {Gargiulo}, {Gomez}, {Marinacci}, {Monachesi}, {Ness},
  {Perez}, {Tissera}, \& {White}}]{Fragkoudietal2020}
{Fragkoudi}, F., {Grand}, R.~J.~J., {Pakmor}, R., {et~al.} 2020, \mnras, 494,
  5936

\bibitem[{{Fragkoudi} {et~al.}(2021){Fragkoudi}, {Grand}, {Pakmor}, {Springel},
  {White}, {Marinacci}, {Gomez}, \& {Navarro}}]{Fragkoudietal2021}
{Fragkoudi}, F., {Grand}, R.~J.~J., {Pakmor}, R., {et~al.} 2021, \aap, 650, L16

\bibitem[{{Ghosh} \& {Jog}(2015)}]{GhoshJog2015}
{Ghosh}, S. \& {Jog}, C.~J. 2015, \mnras, 451, 1350

\bibitem[{{Ghosh} \& {Jog}(2016)}]{GhoshJog2016}
{Ghosh}, S. \& {Jog}, C.~J. 2016, \mnras, 459, 4057

\bibitem[{{Ghosh} \& {Jog}(2018)}]{GhoshJog2018}
{Ghosh}, S. \& {Jog}, C.~J. 2018, \aap, 617, A47

\bibitem[{{Ghosh} \& {Jog}(2022)}]{GhoshJog2021}
{Ghosh}, S. \& {Jog}, C.~J. 2022, \aap, 658, A171

\bibitem[{{Ghosh} {et~al.}(2021){Ghosh}, {Saha}, {Di Matteo}, \&
  {Combes}}]{Ghoshetal2021}
{Ghosh}, S., {Saha}, K., {Di Matteo}, P., \& {Combes}, F. 2021, \mnras, 502,
  3085

\bibitem[{{Gilmore} \& {Reid}(1983)}]{GilmoreandReid1983}
{Gilmore}, G. \& {Reid}, N. 1983, \mnras, 202, 1025

\bibitem[{{Grand} {et~al.}(2016){Grand}, {Springel}, {G{\'o}mez}, {Marinacci},
  {Pakmor}, {Campbell}, \& {Jenkins}}]{Grandetal2016}
{Grand}, R. J.~J., {Springel}, V., {G{\'o}mez}, F.~A., {et~al.} 2016, \mnras,
  459, 199

\bibitem[{{Guo} {et~al.}(2022){Guo}, {Jogee}, {Finkelstein}, {Chen}, {Wise},
  {Bagley}, {Barro}, {Wuyts}, {Kocevski}, {Kartaltepe}, {McGrath}, {Ferguson},
  {Mobasher}, {Giavalisco}, {Lucas}, {Zavala}, {Lotz}, {Grogin},
  {Huertas-Company}, {Vega-Ferrero}, {Hathi}, {Arrabal Haro}, {Dickinson},
  {Koekemoer}, {Papovich}, {Pirzkal}, {Yung}, {Backhaus}, {Bell},
  {Calabr{\`o}}, {Cleri}, {Coogan}, {Cooper}, {Costantin}, {Croton}, {Davis},
  {de la Vega}, {Dekel}, {Franco}, {Gardner}, {Holwerda}, {Hutchison},
  {Pandya}, {P{\'e}rez-Gonz{\'a}lez}, {Ravindranath}, {Rose}, {Trump}, \&
  {Wang}}]{Guoetal2022}
{Guo}, Y., {Jogee}, S., {Finkelstein}, S.~L., {et~al.} 2022, arXiv e-prints,
  arXiv:2210.08658

\bibitem[{{Hamilton-Campos} {et~al.}(2023){Hamilton-Campos}, {Simons},
  {Peeples}, {Snyder}, \& {Heckman}}]{Hamilton-Campos+2023}
{Hamilton-Campos}, K.~A., {Simons}, R.~C., {Peeples}, M.~S., {Snyder}, G.~F.,
  \& {Heckman}, T.~M. 2023, arXiv e-prints, arXiv:2303.04171

\bibitem[{{Haywood} {et~al.}(2013){Haywood}, {Di Matteo}, {Lehnert}, {Katz}, \&
  {G{\'o}mez}}]{Haywoodetal2013}
{Haywood}, M., {Di Matteo}, P., {Lehnert}, M.~D., {Katz}, D., \& {G{\'o}mez},
  A. 2013, \aap, 560, A109

\bibitem[{{Haywood} {et~al.}(2016){Haywood}, {Di Matteo}, {Snaith}, \&
  {Calamida}}]{Haywoodetal2016}
{Haywood}, M., {Di Matteo}, P., {Snaith}, O., \& {Calamida}, A. 2016, \aap,
  593, A82

\bibitem[{{Hernquist} \& {Weinberg}(1992)}]{HernquistWeinberg1992}
{Hernquist}, L. \& {Weinberg}, M.~D. 1992, \apj, 400, 80

\bibitem[{{Hohl}(1971)}]{Hohl1971}
{Hohl}, F. 1971, \apj, 168, 343

\bibitem[{{Hozumi} \& {Hernquist}(2005)}]{HozumiHernqusit2005}
{Hozumi}, S. \& {Hernquist}, L. 2005, \pasj, 57, 719

\bibitem[{{Izquierdo-Villalba} {et~al.}(2022){Izquierdo-Villalba}, {Bonoli},
  {Rosas-Guevara}, {Springel}, {White}, {Zana}, {Dotti}, {Spinoso}, {Bonetti},
  \& {Lupi}}]{Izquierdo-Villalbaetal2022}
{Izquierdo-Villalba}, D., {Bonoli}, S., {Rosas-Guevara}, Y., {et~al.} 2022,
  \mnras, 514, 1006

\bibitem[{{Jean-Baptiste} {et~al.}(2017){Jean-Baptiste}, {Di Matteo},
  {Haywood}, {G{\'o}mez}, {Montuori}, {Combes}, \&
  {Semelin}}]{Jean-Baptisteetal2017}
{Jean-Baptiste}, I., {Di Matteo}, P., {Haywood}, M., {et~al.} 2017, \aap, 604,
  A106

\bibitem[{{Jog}(1996)}]{Jog1996}
{Jog}, C.~J. 1996, \mnras, 278, 209

\bibitem[{{Jog} \& {Solomon}(1984)}]{JogandSolomon1984}
{Jog}, C.~J. \& {Solomon}, P.~M. 1984, \apj, 276, 114

\bibitem[{{Jogee} {et~al.}(2004){Jogee}, {Barazza}, {Rix}, {Shlosman},
  {Barden}, {Wolf}, {Davies}, {Heyer}, {Beckwith}, {Bell}, {Borch}, {Caldwell},
  {Conselice}, {Dahlen}, {H{\"a}ussler}, {Heymans}, {Jahnke}, {Knapen},
  {Laine}, {Lubell}, {Mobasher}, {McIntosh}, {Meisenheimer}, {Peng},
  {Ravindranath}, {Sanchez}, {Somerville}, \& {Wisotzki}}]{Jogeeetal2004}
{Jogee}, S., {Barazza}, F.~D., {Rix}, H.-W., {et~al.} 2004, \apjl, 615, L105

\bibitem[{{Juri{\'c}} {et~al.}(2008){Juri{\'c}}, {Ivezi{\'c}}, {Brooks},
  {Lupton}, {Schlegel}, {Finkbeiner}, {Padmanabhan}, {Bond}, {Sesar},
  {Rockosi}, {Knapp}, {Gunn}, {Sumi}, {Schneider}, {Barentine}, {Brewington},
  {Brinkmann}, {Fukugita}, {Harvanek}, {Kleinman}, {Krzesinski}, {Long},
  {Neilsen}, {Nitta}, {Snedden}, \& {York}}]{Juricetal2008}
{Juri{\'c}}, M., {Ivezi{\'c}}, {\v{Z}}., {Brooks}, A., {et~al.} 2008, \apj,
  673, 864

\bibitem[{{Kasparova} {et~al.}(2016){Kasparova}, {Katkov}, {Chilingarian},
  {Silchenko}, {Moiseev}, \& {Borisov}}]{Kasparovaetal2016}
{Kasparova}, A.~V., {Katkov}, I.~Y., {Chilingarian}, I.~V., {et~al.} 2016,
  \mnras, 460, L89

\bibitem[{{Kim} {et~al.}(2015){Kim}, {Sheth}, {Gadotti}, {Lee}, {Zaritsky},
  {Elmegreen}, {Athanassoula}, {Bosma}, {Holwerda}, {Ho}, {Comer{\'o}n},
  {Knapen}, {Hinz}, {Mu{\~n}oz-Mateos}, {Erroz-Ferrer}, {Buta}, {Kim},
  {Laurikainen}, {Salo}, {Madore}, {Laine}, {Men{\'e}ndez-Delmestre}, {Regan},
  {de Swardt}, {Gil de Paz}, {Seibert}, \& {Mizusawa}}]{Taehyunetal2015}
{Kim}, T., {Sheth}, K., {Gadotti}, D.~A., {et~al.} 2015, \apj, 799, 99

\bibitem[{{Klypin} {et~al.}(2009){Klypin}, {Valenzuela}, {Col{\'\i}n}, \&
  {Quinn}}]{Klypinetal2009}
{Klypin}, A., {Valenzuela}, O., {Col{\'\i}n}, P., \& {Quinn}, T. 2009, \mnras,
  398, 1027

\bibitem[{{Kraljic} {et~al.}(2012){Kraljic}, {Bournaud}, \&
  {Martig}}]{Kraljicetal2012}
{Kraljic}, K., {Bournaud}, F., \& {Martig}, M. 2012, \apj, 757, 60

\bibitem[{{Kruijssen} {et~al.}(2020){Kruijssen}, {Pfeffer}, {Chevance},
  {Bonaca}, {Trujillo-Gomez}, {Bastian}, {Reina-Campos}, {Crain}, \&
  {Hughes}}]{Kruijssenetal2020}
{Kruijssen}, J.~M.~D., {Pfeffer}, J.~L., {Chevance}, M., {et~al.} 2020, \mnras,
  498, 2472

\bibitem[{{Kruk} {et~al.}(2017){Kruk}, {Lintott}, {Simmons}, {Bamford},
  {Cardamone}, {Fortson}, {Hart}, {H{\"a}u{\ss}ler}, {Masters}, {Nichol},
  {Schawinski}, \& {Smethurst}}]{Kruketal2017}
{Kruk}, S.~J., {Lintott}, C.~J., {Simmons}, B.~D., {et~al.} 2017, \mnras, 469,
  3363

\bibitem[{{Martig} {et~al.}(2021){Martig}, {Pinna}, {Falc{\'o}n-Barroso},
  {Gadotti}, {Husemann}, {Minchev}, {Neumann}, {Ruiz-Lara}, \& {van de
  Ven}}]{matigetal2021}
{Martig}, M., {Pinna}, F., {Falc{\'o}n-Barroso}, J., {et~al.} 2021, \mnras,
  508, 2458

\bibitem[{{Martinez-Valpuesta} {et~al.}(2006){Martinez-Valpuesta}, {Shlosman},
  \& {Heller}}]{Martinez-Valpuestaetal2006}
{Martinez-Valpuesta}, I., {Shlosman}, I., \& {Heller}, C. 2006, \apj, 637, 214

\bibitem[{{Masters} {et~al.}(2011){Masters}, {Nichol}, {Hoyle}, {Lintott},
  {Bamford}, {Edmondson}, {Fortson}, {Keel}, {Schawinski}, {Smith}, \&
  {Thomas}}]{Mastersetal2011}
{Masters}, K.~L., {Nichol}, R.~C., {Hoyle}, B., {et~al.} 2011, \mnras, 411,
  2026

\bibitem[{{Melvin} {et~al.}(2014){Melvin}, {Masters}, {Lintott}, {Nichol},
  {Simmons}, {Bamford}, {Casteels}, {Cheung}, {Edmondson}, {Fortson},
  {Schawinski}, {Skibba}, {Smith}, \& {Willett}}]{Melvinetal2014}
{Melvin}, T., {Masters}, K., {Lintott}, C., {et~al.} 2014, \mnras, 438, 2882

\bibitem[{{Mihos} {et~al.}(1997){Mihos}, {McGaugh}, \& {de
  Blok}}]{Mihosetal1997}
{Mihos}, J.~C., {McGaugh}, S.~S., \& {de Blok}, W.~J.~G. 1997, \apjl, 477, L79

\bibitem[{{Miyamoto} \& {Nagai}(1975)}]{MiyamatoandNagai1975}
{Miyamoto}, M. \& {Nagai}, R. 1975, \pasj, 27, 533

\bibitem[{{Nair} \& {Abraham}(2010)}]{NairandAbraham2010}
{Nair}, P.~B. \& {Abraham}, R.~G. 2010, \apjl, 714, L260

\bibitem[{{Ohta} {et~al.}(1990){Ohta}, {Hamabe}, \& {Wakamatsu}}]{Ohtaetal1990}
{Ohta}, K., {Hamabe}, M., \& {Wakamatsu}, K.-I. 1990, \apj, 357, 71

\bibitem[{{Ostriker} \& {Peebles}(1973)}]{OstrikerandPeebles1973}
{Ostriker}, J.~P. \& {Peebles}, P.~J.~E. 1973, \apj, 186, 467

\bibitem[{{Pahwa} \& {Saha}(2018)}]{Pahwa2018}
{Pahwa}, I. \& {Saha}, K. 2018, \mnras, 478, 4657

\bibitem[{{Pfenniger} \& {Norman}(1990)}]{Pfenniger1990}
{Pfenniger}, D. \& {Norman}, C. 1990, \apj, 363, 391

\bibitem[{{Pinna} {et~al.}(2019{\natexlab{a}}){Pinna}, {Falc{\'o}n-Barroso},
  {Martig}, {Coccato}, {Corsini}, {de Zeeuw}, {Gadotti}, {Iodice}, {Leaman},
  {Lyubenova}, {Mart{\'\i}n-Navarro}, {Morelli}, {Sarzi}, {van de Ven},
  {Viaene}, \& {McDermid}}]{Pinnaetal2019b}
{Pinna}, F., {Falc{\'o}n-Barroso}, J., {Martig}, M., {et~al.}
  2019{\natexlab{a}}, \aap, 625, A95

\bibitem[{{Pinna} {et~al.}(2018){Pinna}, {Falc{\'o}n-Barroso}, {Martig},
  {Mart{\'\i}nez-Valpuesta}, {M{\'e}ndez-Abreu}, {van de Ven}, {Leaman}, \&
  {Lyubenova}}]{Pinnaetal2018}
{Pinna}, F., {Falc{\'o}n-Barroso}, J., {Martig}, M., {et~al.} 2018, \mnras,
  475, 2697

\bibitem[{{Pinna} {et~al.}(2019{\natexlab{b}}){Pinna}, {Falc{\'o}n-Barroso},
  {Martig}, {Sarzi}, {Coccato}, {Iodice}, {Corsini}, {de Zeeuw}, {Gadotti},
  {Leaman}, {Lyubenova}, {McDermid}, {Minchev}, {Morelli}, {van de Ven}, \&
  {Viaene}}]{Pinnaetal2019a}
{Pinna}, F., {Falc{\'o}n-Barroso}, J., {Martig}, M., {et~al.}
  2019{\natexlab{b}}, \aap, 623, A19

\bibitem[{{Plummer}(1911)}]{Plummer1911}
{Plummer}, H.~C. 1911, \mnras, 71, 460

\bibitem[{{Pohlen} {et~al.}(2004){Pohlen}, {Balcells}, {L{\"u}tticke}, \&
  {Dettmar}}]{Pohlenetal2004}
{Pohlen}, M., {Balcells}, M., {L{\"u}tticke}, R., \& {Dettmar}, R.~J. 2004,
  \aap, 422, 465

\bibitem[{{Press} {et~al.}(1986){Press}, {Flannery}, \&
  {Teukolsky}}]{Pressetal1986}
{Press}, W.~H., {Flannery}, B.~P., \& {Teukolsky}, S.~A. 1986, {Numerical
  recipes. The art of scientific computing}

\bibitem[{{Rodionov} {et~al.}(2009){Rodionov}, {Athanassoula}, \&
  {Sotnikova}}]{Rodionovetal2009}
{Rodionov}, S.~A., {Athanassoula}, E., \& {Sotnikova}, N.~Y. 2009, \mnras, 392,
  904

\bibitem[{{Romeo} {et~al.}(2022){Romeo}, {Agertz}, \& {Renaud}}]{Romeoetal2022}
{Romeo}, A.~B., {Agertz}, O., \& {Renaud}, F. 2022, arXiv e-prints,
  arXiv:2204.02695

\bibitem[{{Rosas-Guevara} {et~al.}(2022){Rosas-Guevara}, {Bonoli}, {Dotti},
  {Izquierdo-Villalba}, {Lupi}, {Zana}, {Bonetti}, {Nelson}, {Springel},
  {Hernquist}, \& {Vogelsberger}}]{Rosas-Guevaraetal2022}
{Rosas-Guevara}, Y., {Bonoli}, S., {Dotti}, M., {et~al.} 2022, \mnras, 512,
  5339

\bibitem[{{Saha}(2014)}]{Saha2014}
{Saha}, K. 2014, arXiv e-prints, arXiv:1403.1711

\bibitem[{{Saha} \& {Elmegreen}(2018)}]{SahaElmegreen2018}
{Saha}, K. \& {Elmegreen}, B. 2018, \apj, 858, 24

\bibitem[{{Saha} {et~al.}(2018){Saha}, {Graham}, \&
  {Rodr{\'\i}guez-Herranz}}]{Sahaetal2018}
{Saha}, K., {Graham}, A.~W., \& {Rodr{\'\i}guez-Herranz}, I. 2018, \apj, 852,
  133

\bibitem[{{Saha} \& {Naab}(2013)}]{SahaNaab2013}
{Saha}, K. \& {Naab}, T. 2013, \mnras, 434, 1287

\bibitem[{{Saha} {et~al.}(2010){Saha}, {Tseng}, \& {Taam}}]{Sahaetal2010}
{Saha}, K., {Tseng}, Y.-H., \& {Taam}, R.~E. 2010, \apj, 721, 1878

\bibitem[{{Scott} {et~al.}(2021){Scott}, {van de Sande}, {Sharma},
  {Bland-Hawthorn}, {Freeman}, {Gerhard}, {Hayden}, \&
  {McDermid}}]{Scottetal2021}
{Scott}, N., {van de Sande}, J., {Sharma}, S., {et~al.} 2021, \apjl, 913, L11

\bibitem[{{Sellwood} \& {Carlberg}(1984)}]{SellwoodCarlberg1984}
{Sellwood}, J.~A. \& {Carlberg}, R.~G. 1984, \apj, 282, 61

\bibitem[{{Sellwood} \& {Debattista}(2006)}]{SellwoodandDebattista2006}
{Sellwood}, J.~A. \& {Debattista}, V.~P. 2006, \apj, 639, 868

\bibitem[{{Semelin} \& {Combes}(2002)}]{SemelinandCombes2002}
{Semelin}, B. \& {Combes}, F. 2002, \aap, 388, 826

\bibitem[{{Shen} \& {Sellwood}(2004)}]{ShenSellwood2004}
{Shen}, J. \& {Sellwood}, J.~A. 2004, \apj, 604, 614

\bibitem[{{Sheth} {et~al.}(2008){Sheth}, {Elmegreen}, {Elmegreen}, {Capak},
  {Abraham}, {Athanassoula}, {Ellis}, {Mobasher}, {Salvato}, {Schinnerer},
  {Scoville}, {Spalsbury}, {Strubbe}, {Carollo}, {Rich}, \&
  {West}}]{Shethetal2008}
{Sheth}, K., {Elmegreen}, D.~M., {Elmegreen}, B.~G., {et~al.} 2008, \apj, 675,
  1141

\bibitem[{{Simmons} {et~al.}(2014){Simmons}, {Melvin}, {Lintott}, {Masters},
  {Willett}, {Keel}, {Smethurst}, {Cheung}, {Nichol}, {Schawinski},
  {Rutkowski}, {Kartaltepe}, {Bell}, {Casteels}, {Conselice}, {Almaini},
  {Ferguson}, {Fortson}, {Hartley}, {Kocevski}, {Koekemoer}, {McIntosh},
  {Mortlock}, {Newman}, {Ownsworth}, {Bamford}, {Dahlen}, {Faber},
  {Finkelstein}, {Fontana}, {Galametz}, {Grogin}, {Gr{\"u}tzbauch}, {Guo},
  {H{\"a}u{\ss}ler}, {Jek}, {Kaviraj}, {Lucas}, {Peth}, {Salvato}, {Wiklind},
  \& {Wuyts}}]{Simmonsetal2014}
{Simmons}, B.~D., {Melvin}, T., {Lintott}, C., {et~al.} 2014, \mnras, 445, 3466

\bibitem[{{Snaith} {et~al.}(2015){Snaith}, {Haywood}, {Di Matteo}, {Lehnert},
  {Combes}, {Katz}, \& {G{\'o}mez}}]{Snaithetal2015}
{Snaith}, O., {Haywood}, M., {Di Matteo}, P., {et~al.} 2015, \aap, 578, A87

\bibitem[{{Snaith} {et~al.}(2014){Snaith}, {Haywood}, {Di Matteo}, {Lehnert},
  {Combes}, {Katz}, \& {G{\'o}mez}}]{Snaithetal2014}
{Snaith}, O.~N., {Haywood}, M., {Di Matteo}, P., {et~al.} 2014, \apjl, 781, L31

\bibitem[{{Toomre}(1964)}]{Toomre1964}
{Toomre}, A. 1964, \apj, 139, 1217

\bibitem[{{Tremaine} \& {Weinberg}(1984)}]{TremaineWeinberg1984}
{Tremaine}, S. \& {Weinberg}, M.~D. 1984, \mnras, 209, 729

\bibitem[{{Tsikoudi}(1979)}]{Tsikoudi1979}
{Tsikoudi}, V. 1979, \apj, 234, 842

\bibitem[{{van der Kruit} \& {Freeman}(2011)}]{vanderKruit2011}
{van der Kruit}, P.~C. \& {Freeman}, K.~C. 2011, \araa, 49, 301

\bibitem[{{Vieira} {et~al.}(2022){Vieira}, {Carraro}, {Korchagin}, {Lutsenko},
  {Girard}, \& {van Altena}}]{Vieiraetal2022}
{Vieira}, K., {Carraro}, G., {Korchagin}, V., {et~al.} 2022, \apj, 932, 28

\bibitem[{{Vynatheya} {et~al.}(2021){Vynatheya}, {Saha}, \&
  {Ghosh}}]{Vynatheyaetal2021}
{Vynatheya}, P., {Saha}, K., \& {Ghosh}, S. 2021, arXiv e-prints,
  arXiv:2105.03183

\bibitem[{{Whyte} {et~al.}(2002){Whyte}, {Abraham}, {Merrifield}, {Eskridge},
  {Frogel}, \& {Pogge}}]{Whyteetal2002}
{Whyte}, L.~F., {Abraham}, R.~G., {Merrifield}, M.~R., {et~al.} 2002, \mnras,
  336, 1281

\bibitem[{{Wozniak} {et~al.}(1995){Wozniak}, {Friedli}, {Martinet}, {Martin},
  \& {Bratschi}}]{Wozniak1995}
{Wozniak}, H., {Friedli}, D., {Martinet}, L., {Martin}, P., \& {Bratschi}, P.
  1995, \aaps, 111, 115

\bibitem[{{Wozniak} \& {Pierce}(1991)}]{Wozniak1991}
{Wozniak}, H. \& {Pierce}, M.~J. 1991, \aaps, 88, 325

\bibitem[{{Yoachim} \& {Dalcanton}(2006)}]{Yoachim2006}
{Yoachim}, P. \& {Dalcanton}, J.~J. 2006, \aj, 131, 226

\end{thebibliography}

\begin{appendix}

\section{$R_{\rm bar}$ measurement for the thin+thick models}
\label{appen:rbar_measurement}

In past literatures, both observationally and from simulations, a wide variety of techniques has been employed to measure the bar length, $R_{\rm bar}$. This includes isophotal fitting with ellipse \cite[e.g.,][]{Wozniak1991,Wozniak1995}, Fourier analysis of the azimuthal luminosity profiles \citep[e.g.,][]{Ohtaetal1990}, measuring the location of the maximum of bar-- inter-bar luminosity ratio \citep{Aguerrietal2000}, as well as by finding the deviation of the phase angle by a certain amount ($5-10 \degrees$), measuring the location where the $A_2/A_0$ ($m=2$ Fourier coefficient) drops to a certain fraction of its peak value.
\par
\begin{figure}
\includegraphics[width=\linewidth]{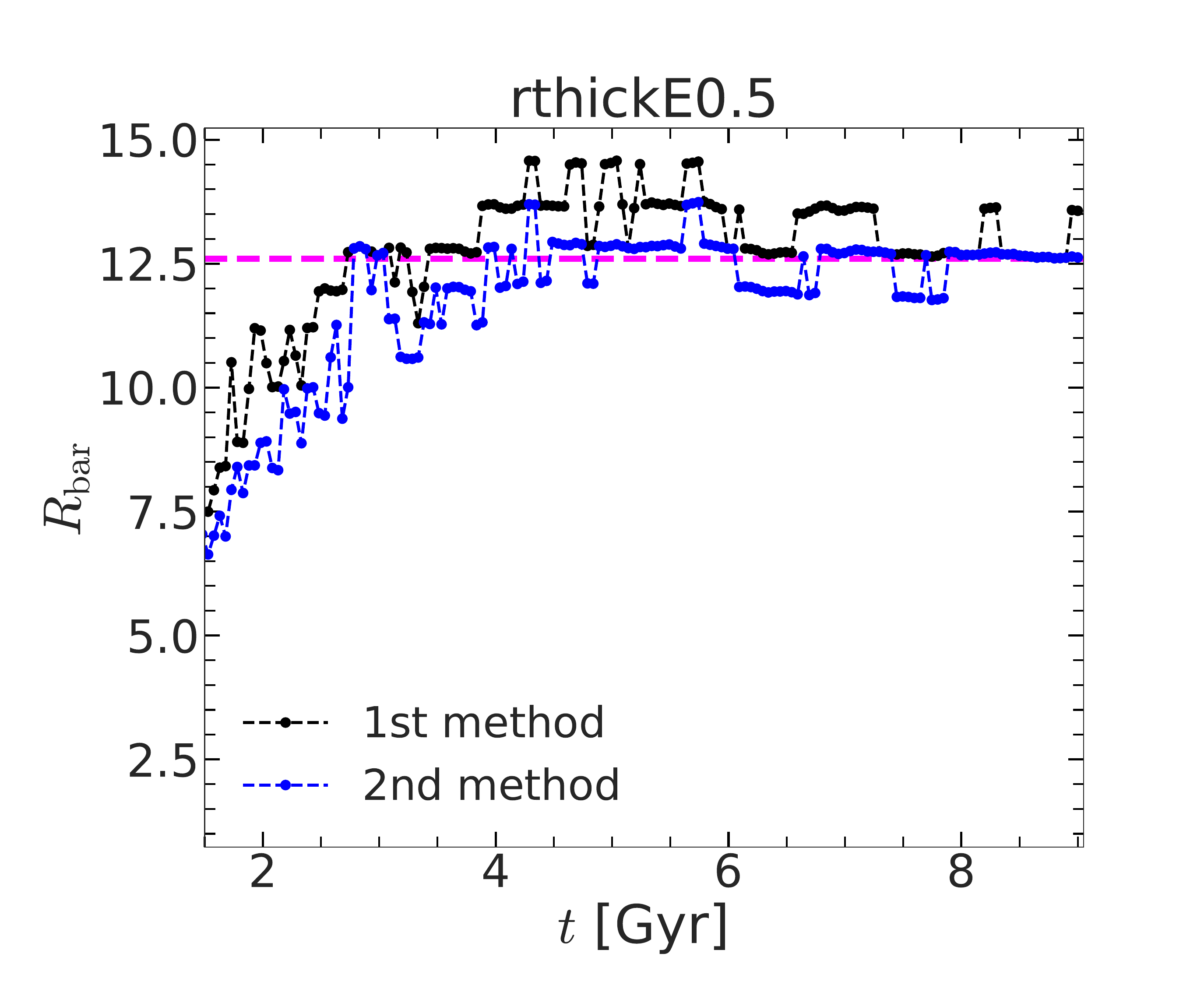}
\caption{Temporal evolution of the bar length, $R_{\rm bar}$, measured by two methods adopted here (see text for details),  for the model \textbf{rthickE0.5}. The horizontal magenta line denotes the representative value for the $R_{\rm bar}$ used for this model.}
\label{fig:rbar_measurement}
\end{figure}

Here, we briefly mention how we calculate the $R_{\rm bar}$ in our thin+thick disc models. To achieve that, we make use of two methods. In the first method, we define the $R_{\rm bar}$ as the location where the $A_2/A_0$ value drops to 60 percent of its peak value. In the second method, we define the $R_{\rm bar}$ as the location where the $A_2/A_0$ value drops to 70 percent of its peak value. The corresponding temporal evolution of the $R_{\rm bar}$, calculated for the total (thin+thick) disc particles are shown in Fig.~\ref{fig:rbar_measurement} for the model \textbf{rthickE0.5}. As seen clearly, the values of $R_{\rm bar}$ increases quite significantly, almost by a factor of two or so. However, after $6 \Gyr$ or so, the bar does not grow appreciably, and the values of $R_{\rm bar}$ also saturates, with occasional fluctuations (compare Figs.~\ref{fig:barStrength_evolution} and \ref{fig:rbar_measurement}). Therefore, we take this value as the representative value for the $R_{\rm bar}$ (see the horizontal magenta line in Fig.~\ref{fig:rbar_measurement}). The $R_{\rm bar}$ values for all other thin+thick disc models are determined in this fashion, and they are indicated in Fig.~\ref{fig:density_maps_endstep_allmodels} (see black dashed circles there).

\end{appendix}

\end{document}